\documentclass[10pt,aps,prc,amsmath,amssymb,twocolumn,floatfix,nofootinbib,showpacs,noeprint,superscriptaddress]{revtex4}

\usepackage{graphicx}
\usepackage{color}
\usepackage{graphicx}
\usepackage{amsmath,amssymb,bm,amsfonts}
\usepackage{hyperref}

\graphicspath{{figures/}}

\definecolor{darkred}{rgb}{0.75, 0.0, 0.0}

\newcommand{\beq}{\begin{equation}}
\newcommand{\eeq}{\end{equation}}
\newcommand{\bea}{\begin{eqnarray}}
\newcommand{\eea}{\end{eqnarray}}

\newcommand{\fmi}{\, \text{fm}^{-1}}

\newcommand{\hw}{\ensuremath{\hbar\Omega}}

\newcommand{\kinf}{\ensuremath{k_{\infty}}}

\newcommand{\Einf}{\ensuremath{E_{\infty}}}
\newcommand{\Ainf}{\ensuremath{A_{\infty}}}

\newcommand{\LamUV}{\Lambda_{UV}}
\newcommand{\LA}{L'_0}  
\newcommand{\ANC}{\gamma_\infty}
\newcommand{\Jost}{\mathfrak{f}}

\begin{document}

\begin{titlepage}
  \title{Universal properties of infrared oscillator basis
    extrapolations} 

  \author{S.N.\ More} \email{more.13@osu.edu} \affiliation{Department
    of Physics, The Ohio State University, Columbus, OH 43210}
  
  \author{A.~Ekstr\"om} \affiliation{Department of Physics and Center
    of Mathematics for Applications, University of Oslo, N-0316 Oslo,
    Norway} \affiliation{National Superconducting Cyclotron
    Laboratory, Michigan State University, East Lansing, MI
    48824-1321, USA}

  \author{R.J.\ Furnstahl} \email{furnstahl.1@osu.edu}
  \affiliation{Department of Physics, The Ohio State University,
    Columbus, OH 43210}

  \author{G.~Hagen} \affiliation{Physics Division, Oak Ridge National
    Laboratory, Oak Ridge, Tennessee 37831, USA}
  \affiliation{Department of Physics and Astronomy, University of
    Tennessee, Knoxville, Tennessee 37996, USA} 

  \author{T.~Papenbrock} \email{tpapenbr@utk.edu}
  \affiliation{Department of Physics and Astronomy, University of
    Tennessee, Knoxville, Tennessee 37996, USA} \affiliation{Physics
    Division, Oak Ridge National Laboratory, Oak Ridge, Tennessee
    37831, USA}

\date{\today}

\begin{abstract}
  Recent work has shown that a finite harmonic oscillator basis in
  nuclear many-body calculations effectively imposes a hard-wall
  boundary condition in coordinate space, motivating infrared
  extrapolation formulas for the energy and other observables. Here we
  further refine these formulas by studying two-body models and the
  deuteron. We accurately determine the box size as a function of the
  model space parameters, and compute scattering phase shifts in the
  harmonic oscillator basis.  We show that the energy shift can be
  well approximated in terms of the asymptotic normalization
  coefficient and the bound-state momentum, discuss higher-order
  corrections for weakly bound systems, and illustrate this universal
  property using unitarily equivalent calculations of the deuteron.
\end{abstract}

\smallskip
  \pacs{21.30.-x,05.10.Cc,13.75.Cs}
\maketitle

\end{titlepage}

\section{Introduction}

Harmonic oscillator (HO) basis expansions are widely used in nuclear
structure calculations, but limited computational resources often
require that the basis be truncated before observables are fully
converged.  In such cases, a procedure to extrapolate results to
infinite basis size is needed.  Such schemes have conventionally been
formulated using the basic parameters defining the oscillator space,
namely the maximum number of oscillator quanta $N$ and the frequency
$\Omega$ of the oscillator wave functions.  An alternative approach to
extrapolations is motivated by effective field theory (EFT) and based
instead on explicitly considering the infrared (IR) and ultraviolet
(UV) cutoffs imposed by a finite oscillator
basis~\cite{Coon:2012ab,Furnstahl:2012qg}.  This has recently led to a
theoretically motivated IR correction formula and an empirical UV
correction formula~\cite{Furnstahl:2012qg} in which the basic
extrapolation variables are an effective hard-wall radius $L$ and the
analogous cut-off in momentum, $\LamUV$.  In terms of the oscillator
length $b\equiv\sqrt{\hbar/(m\Omega)}$, rough estimates of these
variables are $L \approx \sqrt{2(N+3/2)} b \equiv L_0$ and $\LamUV
\approx \sqrt{2(N+3/2)}\hbar/b$~\cite{Coon:2012ab,Furnstahl:2012qg}.

The $b$ dependence of $L$ and $\LamUV$ suggests that if the oscillator
length is small enough (i.e., if the oscillator frequency is large
enough), the UV correction will be negligible compared to the IR
correction.  In this domain, an estimate for the energy in the
truncated basis was derived in Ref.~\cite{Furnstahl:2012qg} based on
an effective Dirichlet boundary condition at $L$:
\beq
      E(L) = \Einf + A e^{-2 \kinf L} + \mathcal{O}(e^{-4\kinf L})
      \;,
   \label{eq:IR_scaling1}
\eeq
where $\kinf = \sqrt{-2 m \Einf/\hbar^2}$ is the binding momentum
defined from the separation energy $\Einf$.  Consideration of the
tails of the HO wave functions motivated an improved choice for $L$
given $N$ and $\hw$~\cite{Furnstahl:2012qg}:
\beq
   \LA \approx L_0 + 0.54437\, b\, (L_0/b)^{-1/3}
   \;.
   \label{eq:LA}
\eeq
The extrapolation formula~(\ref{eq:IR_scaling1}) is the leading-order 
correction to the
ground-state energy once UV corrections can be neglected and once $L$
exceeds the radius of the nucleus under
consideration.  Test calculations of few- and many-body nuclei using
$L = \LA$ and with $\Einf$, $A$, and $\kinf$ as fit parameters showed
that the IR correction formula~\eqref{eq:IR_scaling1} can be used in
practice~\cite{Furnstahl:2012qg}.  (Note: The results in
Ref.~\cite{Furnstahl:2012qg} were derived in the laboratory system
with $m$ the particle mass.  Here for convenience we take $m$ to be
the reduced mass $\mu$, which rescales $b$ and $\kinf$ but leaves the
expressions unchanged.)

In the present work we seek a more complete understanding of this
correction formula and to more accurately determine the hard-wall radius
$L$.  While the most useful application of Eq.~\eqref{eq:IR_scaling1}
is to few- or many-body nuclei, we specialize here to the two-particle
case, which we can control and calculate precisely.  
In doing so we gain insight into the
universal features of the IR extrapolation, including its invariance
to phase-shift equivalent potentials and its application to excited
states.  While the coefficient $A$ was previously treated purely as a
fit parameter, we extend the derivation from
Ref.~\cite{Furnstahl:2012qg} to show how it can be expressed in terms
of the observables $\kinf$ and the asymptotic normalization constant
$\ANC$, just as in the related L\"uscher-type formulas developed for
lattice
applications~\cite{Luscher:1985dn,Lee:2010km,Pine:2012zv,Konig:2011ti}.
We examine the approximations leading to Eq.~\eqref{eq:IR_scaling1}
and derive a corrected formula appropriate for weakly bound states,
which is shown to work much better for the deuteron.

Our strategy is to use a range of model potentials for which the
Schr\"odinger equation can be solved
analytically or to any desired precision numerically to broadly test
and illustrate various features, and then turn to the deuteron with a
set of phase-shift equivalent potentials for a real-world example.  In
particular we will consider:
\bea
  V_{\rm sw}(r) &=& -V_0\, \theta(R-r)   \qquad \mbox{[square well]}
  \;,
  \label{eq:Vsw}
  \\
  V_{\rm exp}(r) &=& -V_0\, e^{-(r/R)}  \qquad\ \ \mbox{[exponential]}
  \;,
  \\
  V_{\rm g}(r) &=& -V_0\, e^{-(r/R)^2}  \qquad\ \mbox{[Gaussian]}
  \;,
  \label{eq:Vg}
  \\
  V_{\rm q}(r) &=& -V_0\, e^{-(r/R)^4} \qquad\ \mbox{[quartic]}
  \;,
  \label{eq:Vq}
\eea
where for each of the models we work in units with $\hbar = 1$, reduced mass
$\mu=1$, and express all lengths in units of $R$ and all energies in units
of $\hbar^2/\mu R^2$.  For the
realistic potential we use the Entem-Machleidt 500\,MeV chiral EFT
N$^3$LO potential~\cite{Entem:2003ft} and unitarily evolve it with the
similarity renormalization group (SRG).  These potentials provide a
diverse set of tests for universal properties.  Because we can go to
very high \hw\ and $N$ for the two-particle bound states (and
therefore large $\LamUV$), it is possible to always ensure that UV
corrections are negligible.

In Section~\ref{sec:HO_vs_Dirichlet} we determine a more accurate
value for $L$ than $\LA$ and show that the theoretically founded
exponential form of the extrapolation is favored over Gaussian or
power-law alternatives in practical applications. The accurate
determination of the box radius $L$ also allows us to compute
scattering phase shifts directly in the oscillator basis. The
derivation of the exponential form from Ref.~\cite{Furnstahl:2012qg}
is extended in Section~\ref{sec:universality} to show that it depends
only on observable quantities, and is therefore independent of the
potential and has the same form for excited states.  These formal
conclusions are tested with model potentials and the deuteron with a
realistic potential in Section~\ref{sec:tests}.  In
Section~\ref{sec:summary} we summarize our conclusions and discuss the
implications for applications to larger nuclei.

\section{Spatial cutoff from  HO basis truncations}
 \label{sec:HO_vs_Dirichlet}
 
 In this Section, we determine the spatial extent of a finite HO
 basis. We start with empirical considerations before presenting an
 analytical understanding. Finally, we use the knowledge of the
 spatial extent to compute phase shifts and demonstrate
 that the theoretically founded exponential extrapolation law can be
 distinguished from other empirical choices.

\subsection{Empirical determination of $L$}

\begin{figure}
 \includegraphics[width=0.95\columnwidth]{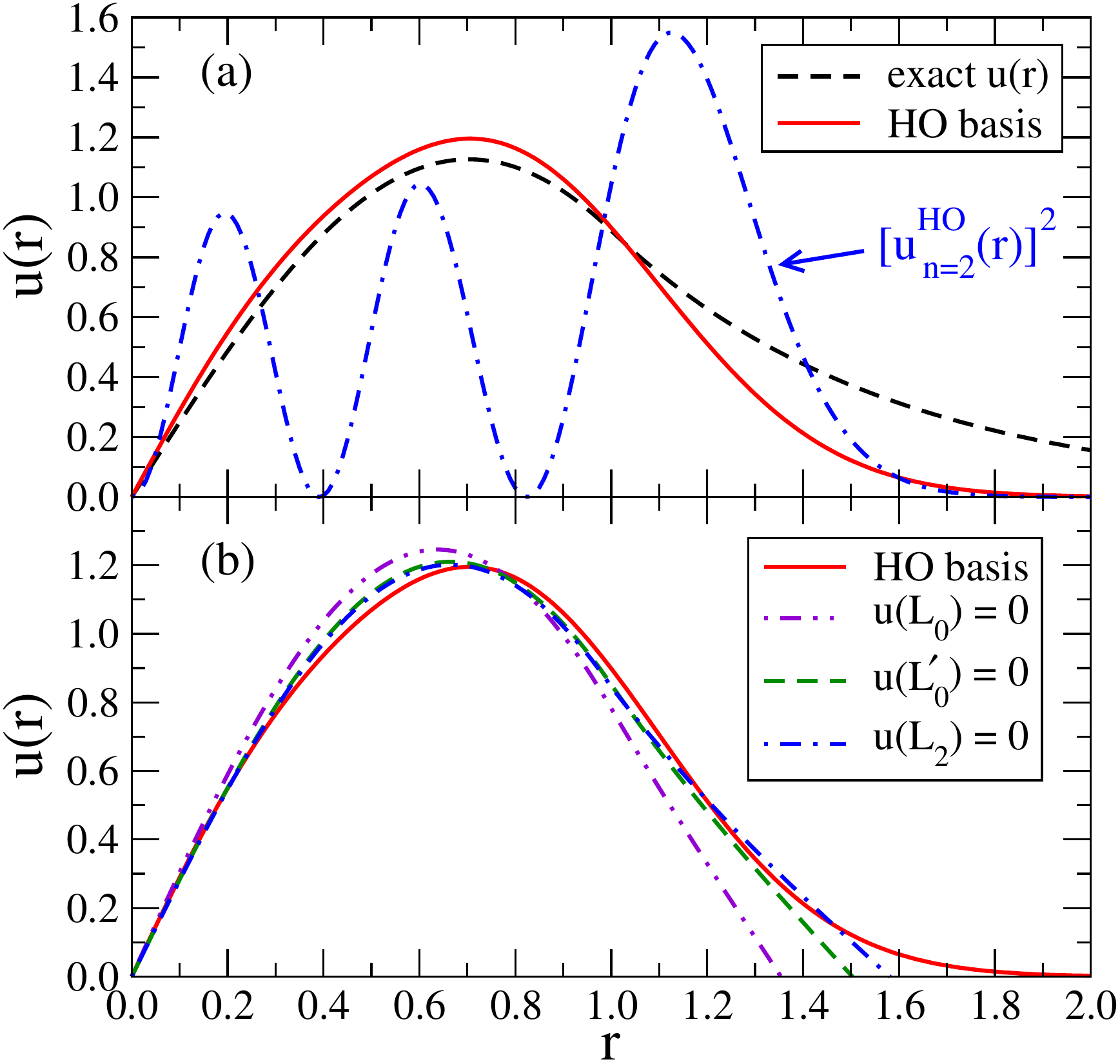}
 \caption{(color online) (a) The exact radial wave
   function (dashed) for a square well Eq.~\eqref{eq:Vsw} with depth
   $V_0=4$ (and $\hbar = \mu = R=1$) is compared to the wave function obtained
   from an HO basis truncated at $N=4$ with $\hw=6$ (solid).
   The spatial extent of
   the wave function obtained from the HO basis truncation is dictated by
   the square of HO wave function for the highest radial quantum
   number (dot-dashed). (b) The wave functions obtained from imposing a Dirichlet
   boundary condition at $L_0$, $\LA$ and $L_2$ are compared to the wave function in 
   truncated HO basis.}
 \label{fig:wavefunctions}
\end{figure}
 
The derivation of the IR correction formula Eq.~\eqref{eq:IR_scaling1}
in Ref.~\cite{Furnstahl:2012qg} starts from the observation that a
truncated harmonic oscillator (HO) basis effectively acts  at low energies
to impose a
hard-wall boundary condition in coordinate space.  In
Fig.~\ref{fig:wavefunctions} we can see how this happens for a
representative model case, a square well potential Eq.~\eqref{eq:Vsw}
with $s$-wave radial wave functions.  In the top panel, the exact
ground-state radial wave function (dashed) is compared to the solution
in an oscillator basis truncated at $N=4$ determined by
diagonalization (solid).  The truncated basis cuts off the tail of the
exact wave function because the individual basis wave functions have a
radial extent that depends on \hw\ (from the Gaussian part) and on the
largest power of $r$ (from the polynomial part).  The latter is given
by $N = 2n + l$.  With $N = 4$ and $l=0$, this means that $n=2$ gives
the largest power.

The cutoff will then be determined by the $n=2$ oscillator wave
function, $u_{n=2}^{\rm HO}(r)$, whose square (which is the relevant
quantity) is also plotted in the top panel (dot-dashed).  It is
evident that the tail of the wave function in the truncated basis is
fixed by this squared wave function.  The premise of
Ref.~\cite{Furnstahl:2012qg} was that this cutoff is well modeled by a
hard-wall (Dirichlet) boundary condition at $r=L$.  If so, the
question remains how best to \emph{quantitatively} determine $L$ given
$N$ and $\hw$. Before we present an analytical derivation of this quantity in the next Subsection, we compare empirically $L'_0$ from Eq.~(\ref{eq:LA}) and 
\beq
   L_i \equiv \sqrt{2(N + 3/2 + i)} b
   \label{eq:L_i}
\eeq
with integer $i$, which includes $L_0$ as a special case. In the
bottom panel of Fig.~\ref{fig:wavefunctions} we show the wave functions
for several possible choices for $L$.  
$L_0$ corresponds to choosing the classical
turning point (i.e. the half-height point of the tail of
$[u^{HO}_{n=2}(r)]^2$); it is manifestly too small.  Using $L'_0$,
which is the linear extrapolation from the slope at the half-height
point, gives an improved estimate.  However, choosing $i=2$ (i.e.,
using $L = L_2 = \sqrt{2(N +3/2+2)}b$) is found to be the best
choice in almost all examples.

\begin{figure}[tbh-]
\includegraphics[width=0.95\columnwidth]{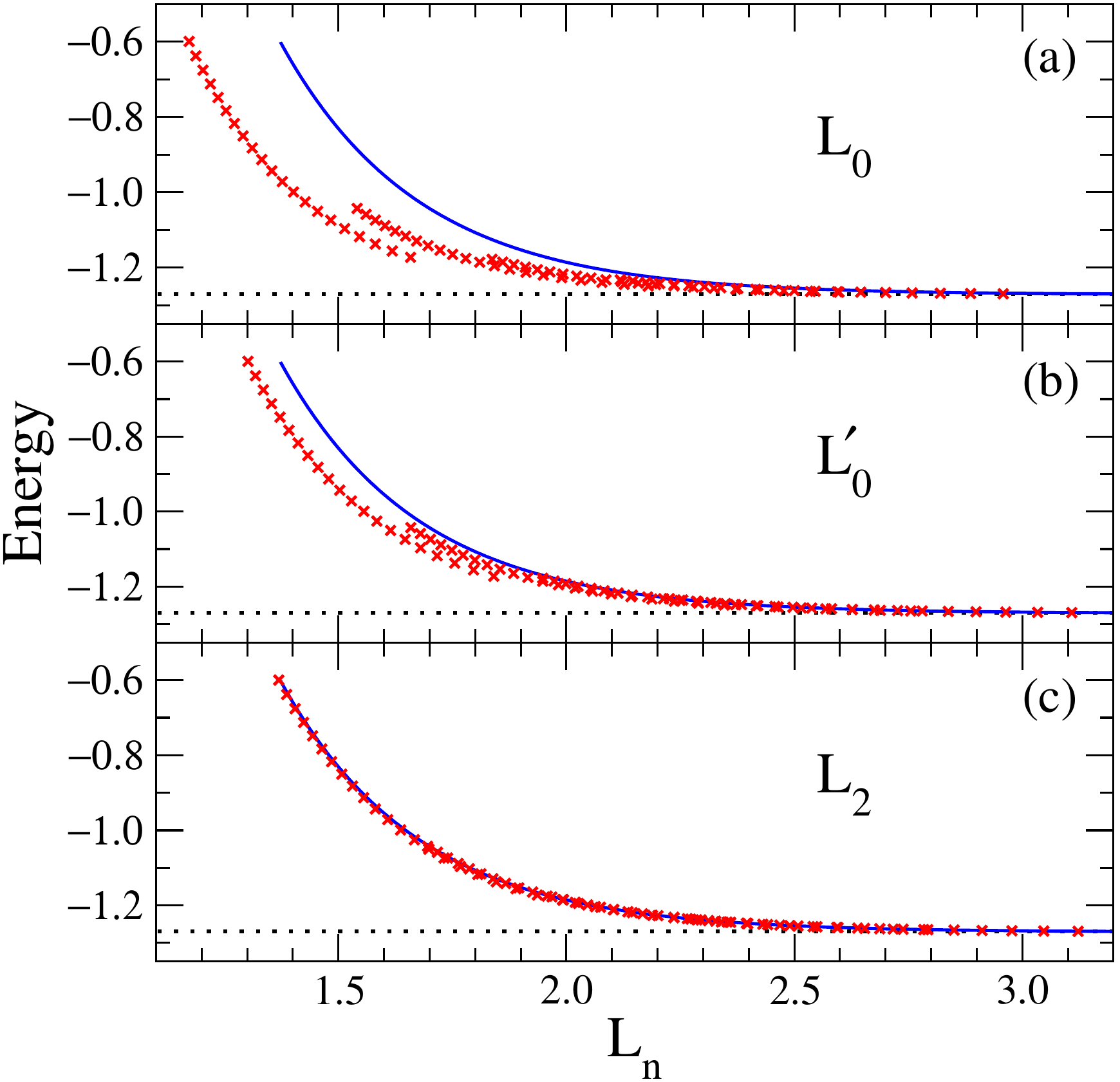}
\caption{(color online) Ground-state energies versus (a) $L_0$,
  (b) $\LA$, and (c) $L_2$ for a Gaussian potential well
  Eq.~\eqref{eq:Vg} with $V_0=5$ (and $\hbar = \mu = R=1$). 
  The crosses are the
  energies from HO basis truncation. 
  The energies obtained by
  numerically solving the Schr{\"o}dinger equation with a Dirichlet
  boundary condition at $L$ lie on the solid line. 
  The horizontal dotted lines mark
  the exact energy, $\Einf=-1.27$.}
\label{fig:spatial_cutoff_vs_HO_gauss}
\end{figure}

The most direct illustration of this conclusion comes from the
bound-state energies.  In the example in Fig.~\ref{fig:wavefunctions},
the exact energy (in dimensionless units) is $-1.51$ while the
result for the basis truncated at $N=4$ is $-1.33$, which is therefore what we
hope to reproduce.  With $L_0$, the energy is $-0.97$, with $\LA$ it
is $-1.21$, and with $L_2$ it is $-1.29$.  While this is only one
example of a model problem, we have found that $L_2$ always gives a
better energy estimate than $\LA$ (and $L_3$ is almost always worse).

\begin{figure}[tbh-]
\includegraphics[width=0.95\columnwidth]{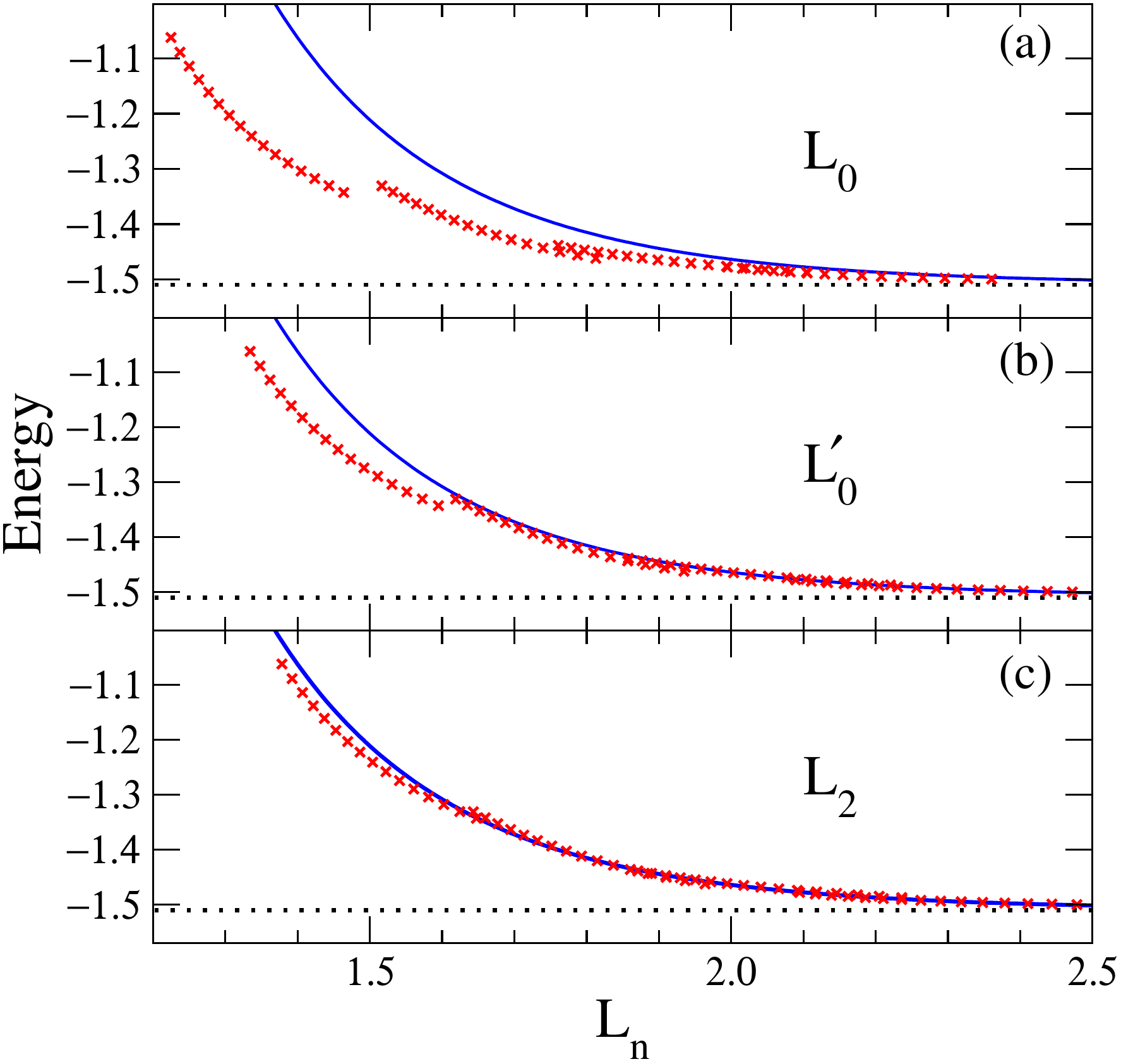}
\caption{(color online) Ground-state energies versus (a) $L_0$,
  (b) $\LA$, and (c) $L_2$ for a square well potential well
  Eq.~\eqref{eq:Vsw} with $V_0=4$ (and $\hbar = \mu = R=1$).
  The crosses are the
  energies from HO basis truncation. 
  The energies obtained by
  numerically solving the Schr{\"o}dinger equation with a Dirichlet
  boundary condition at $L$ lie on the solid line.  
  The horizontal dotted lines mark
  the exact energy, $\Einf=-1.51$.}
\label{fig:spatial_cutoff_vs_HO_square_well}
\end{figure}

\begin{figure}[tbh-]
\includegraphics[width=0.95\columnwidth]{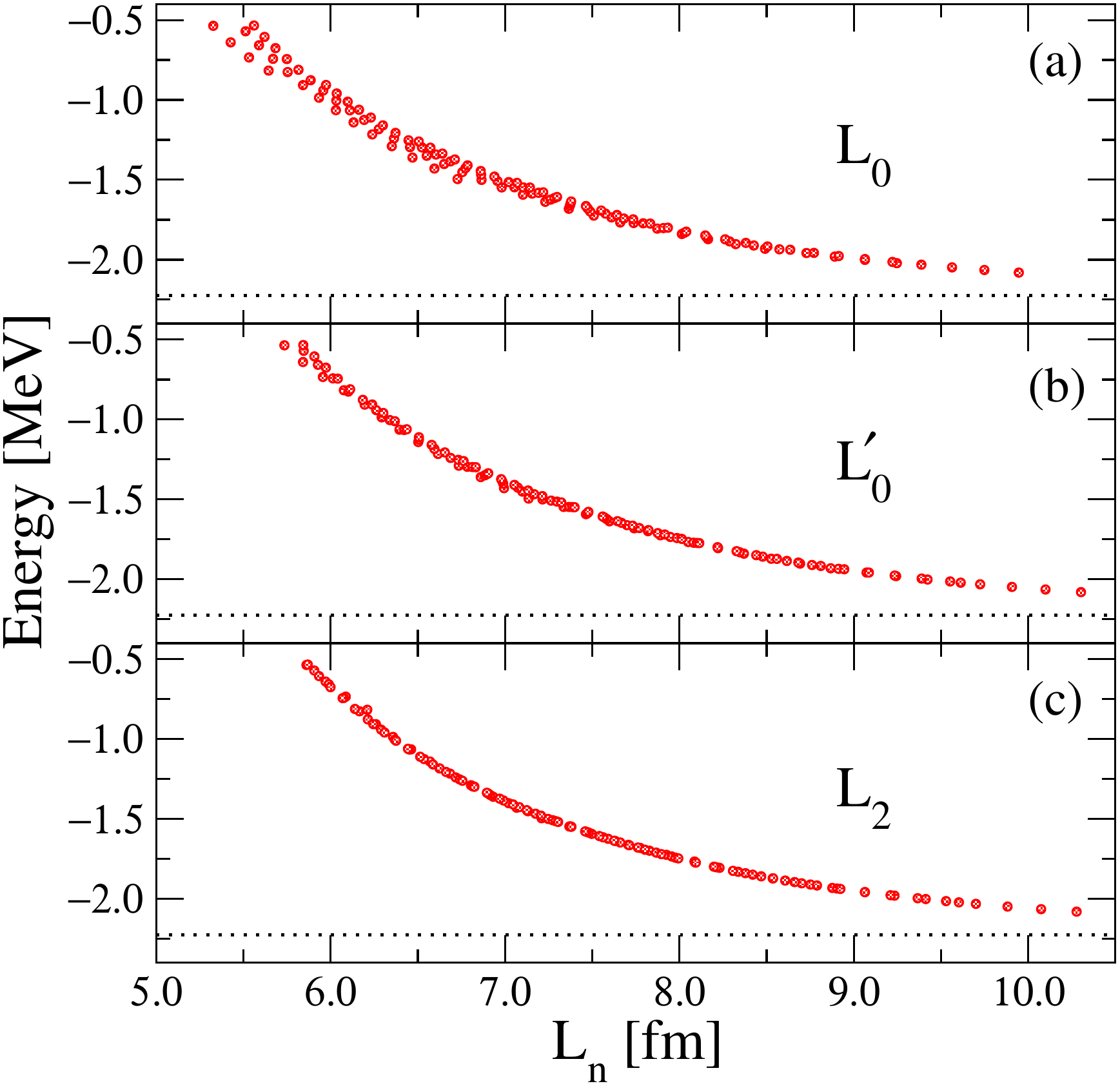}
\caption{(color online) Ground-state energies versus (a) $L_0$,
  (b) $\LA$, and (c) $L_2$ for the Entem-Machleidt 500\,MeV
  N$^3$LO potential~\cite{Entem:2003ft}.  The horizontal dotted lines
  mark the exact energy, $\Einf = -2.2246\,\mbox{MeV}$. }
\label{fig:spatial_cutoff_vs_HO_deuteron}
\end{figure}

Another signature that demonstrates the suitability of $L_2$ is that
points from many different $\hw$ and $N$ values all lie on the same
curve. Figures.~\ref{fig:spatial_cutoff_vs_HO_gauss} and
\ref{fig:spatial_cutoff_vs_HO_square_well} show the energies from a
wide range of HO truncations for $L_0$, $\LA$ and $L_2$ for the
Gaussian well and the square well potential, respectively.  The
energies for different $\hbar\Omega$ and $N$ lie on the same smooth
and unbroken curve if we use $L_2$ but not with the other choices. For
$L=L_0$ and $L=\LA$, one finds that sets of points with different
$\hbar\Omega$ but same $N$ fall on smooth, $N$-dependent curves.  For
the square well, there are small discontinuities visible even for
$L=L_2$. At the square well radius, the wave function's second
derivative is not smooth, and this is difficult to approximate with a
finite set of oscillator functions. This lack of UV convergence is
likely the origin of the very small discontinuities.
As a further test, we solve the Schr{\"o}dinger equation with a
vanishing Dirichlet boundary condition (solid lines in
Figs.~\ref{fig:spatial_cutoff_vs_HO_gauss} and
\ref{fig:spatial_cutoff_vs_HO_square_well}) and compare to the
energies obtained from the HO truncations (crosses).  The finite
oscillator basis energies are well approximated by a Dirichlet
boundary condition with a mapping from the oscillator $\hbar\Omega$
and $N$ to an equivalent length given by $L_2$.  Note that for large
$N$, the differences between $L_0$, $\LA$ and $L_2$ may be smaller
than other uncertainties involved in nuclear calculations, but for
practical calculations one will want to use small $N$ results, where
these considerations are very relevant.

\begin{figure}[tbh-]
\includegraphics[width=0.95\columnwidth]{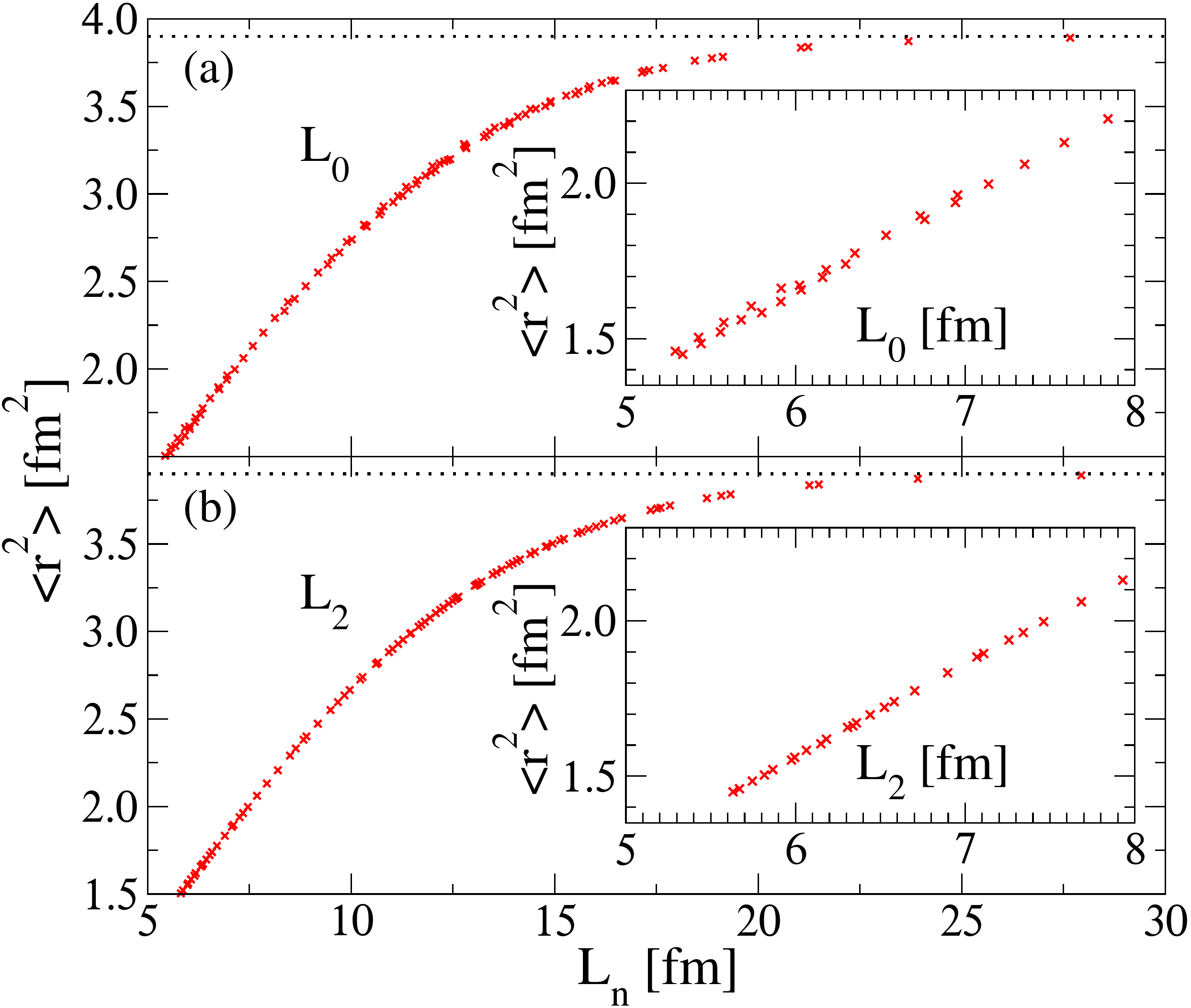}
\caption{(color online) Deuteron radius squared versus (a) $L_0$ and
  (b) $L_2$ for the Entem-Machleidt 500\,MeV N$^3$LO
  potential~\cite{Entem:2003ft}.  The horizontal dotted lines mark the
  exact radius squared, $r^2_{\infty} = 3.9006~{\rm fm}^2$. The insets
  show a magnification of data at smaller lengths $L_n$.}
\label{fig:deuteron_radii}
\end{figure}

These results from model calculations are consistent with those from
realistic potentials applied to the deuteron.  To illustrate this, we
use the N$^3$LO 500\,MeV potential of Entem and
Machleidt~\cite{Entem:2003ft}.  We generate results in an HO basis
with \hw\ ranging from $1$ to $100\,\mbox{MeV}$ and $N$ from $4$ to
$100$ (in steps of 4 to avoid HO artifacts for the
deuteron~\cite{Bogner:2007rx}).  We then restrict the data to where UV
corrections are negligible (see Section~\ref{subsec:SRG}).
Figure~\ref{fig:spatial_cutoff_vs_HO_deuteron} shows that the
criterion of a continuous curve with the smallest spread of points
clearly favors $L_2$. Similar comments apply to the computation
of the radius. Figure~\ref{fig:deuteron_radii} shows that the
numerical results for the squared radius, when plotted as a function
of $L_2$ (but not as a function of $L_0$), fall on a continuous curve
with minimal spread.

\subsection{Analytical derivation of $L_2$}
\label{subsec:analytic}

\begin{figure}[tbh-]
\includegraphics[width=0.95\columnwidth]{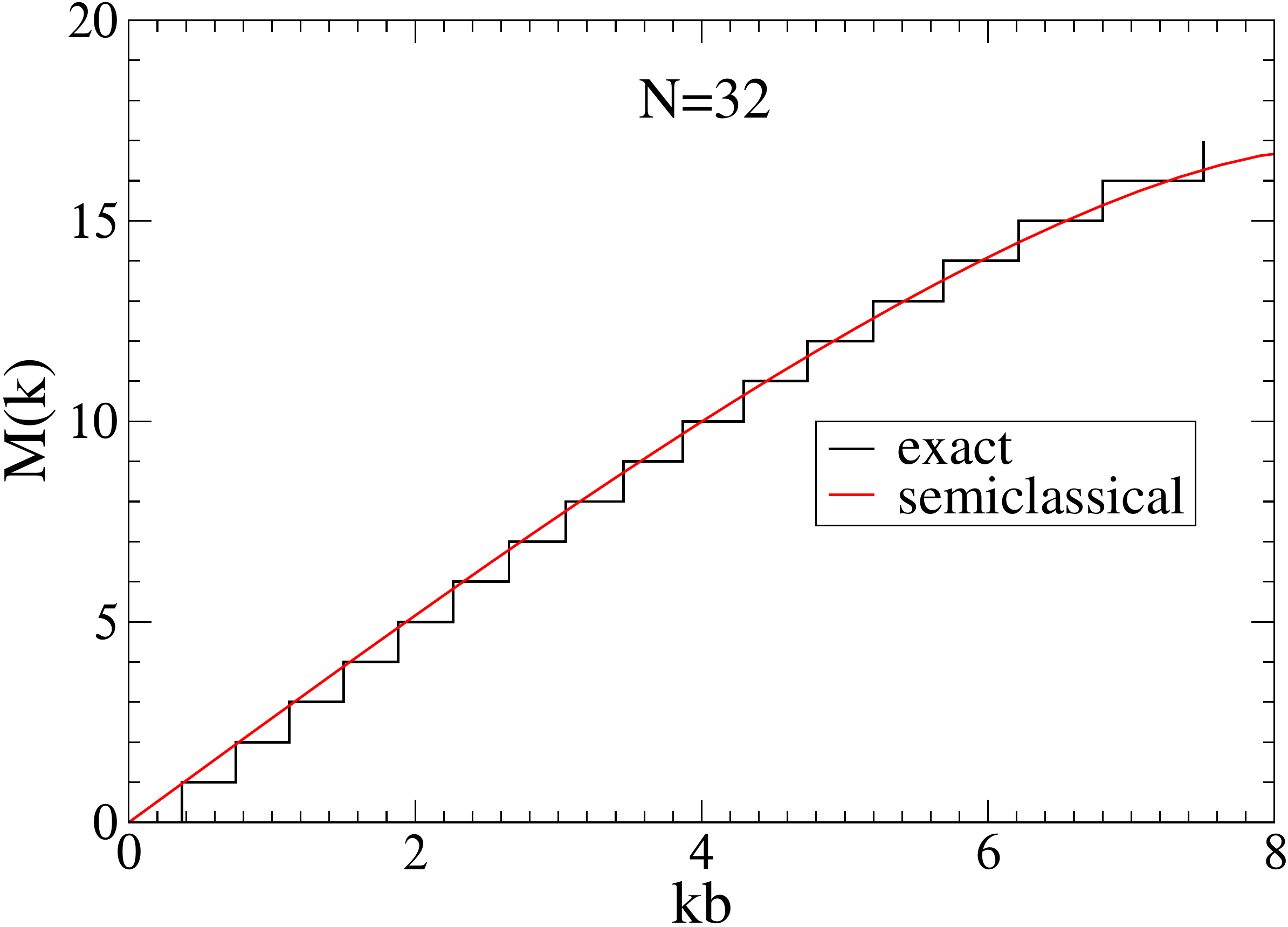}
\caption{(color online) The staircase function of the $s$ states of
  the operator $p^2$ in a finite oscillator basis with $N=32$ (black)
  compared to its semiclassical estimate (smooth red curve). $M(k)$
  denotes the number of states of the operator $p^2$ with eigenvalues
  $p^2\le\hbar^2 k^2$.}
\label{fig:staircase}
\end{figure}

Naturally, the squared momentum operator $p^2$ is the key for
understanding the IR properties of the harmonic oscillator basis.  Let
us start with the spectrum of $p^2$ in the oscillator
basis. In a finite basis with energies up to $E=(N+3/2)\hbar\Omega$,
the operator $p^2$ must be viewed as
$p^2\Theta(E-p^2/(2m)-(m/2)\Omega^2 r^2)$, where $\Theta$ denotes the
unit step function. Let us compute the number $M(k)$ of $s$-wave
states up to a momentum $k$ as a first step. We find
\bea 
  M(k)&=& {\rm Tr} \left[\Theta\left(\hbar^2k^2-p^2\right)
   \Theta\left(E-{p^2\over 2m}-{m\over 2}\Omega^2 r^2\right) \right] 
  \nonumber\\
  &\approx&{1\over 2\pi\hbar} \int\limits_{-\hbar k}^{\hbar k}\! dp
    \int\limits_0^\infty\! dr \, 
   \Theta\left(\hbar^2k^2-p^2\right) \nonumber\\
   &&\null\times \Theta\left(E-{p^2\over 2m}-{m\over 2}\Omega^2 r^2\right) 
   \; .
\eea
Here, we apply the semiclassical approximation and write the trace
as a phase-space integral.  We assume $\hbar^2k^2/(2m)\le
E$, perform the integrations and use $E/(\hbar\Omega)= N+3/2$. This
yields
\bea
  M(k)&=& {bk\over 2\pi}\sqrt{2N+3-b^2k^2} \nonumber\\
  && \null +{N+3/2\over \pi}\arcsin{bk\over\sqrt{2N+3}} 
  \;,
  \label{Mstair}
\eea
where $b$ is the oscillator length.
Figure~\ref{fig:staircase} shows a comparison between the quantum
mechanical staircase function and the semiclassical
estimate~(\ref{Mstair}) for $N=32$. For sufficiently small values of
$kb\ll\sqrt{2N}$, the number of $s$-wave momentum eigenstates grows
linearly, and inspection of Eq.~(\ref{Mstair}) shows that the slope at
the origin is $L_0/\pi$ semiclassically. The linear growth of the
number of eigenstates of $p^2$ with $k$ clearly demonstrate that --- at
not too large values of $kb$ --- the spectrum of $p^2$ in the oscillator
basis is indistinguishable from the spectrum of $p^2$ in a spherical
box. For the determination of the box radius $L$, we note that the
lowest eigenvalue of $p^2$ is $(\pi/L)^2$.

In what follows, we analytically compute the smallest eigenvalue
$\kappa^2_{\rm min}$ of $p^2$ in a finite oscillator basis and will
see that $\kappa_{\rm min} = \pi/L_2$. In the remainder of this
Subsection, we set the oscillator length to one. We focus on $s$-waves
and thus consider wave functions that are regular at the origin, i.e.
the radial wave functions are identical to the odd wave functions of
the one-dimensional harmonic oscillator.

The localized eigenfunction of the operator $p^2$ with smallest
eigenvalue $\kappa^2$ is
\bea
\label{eigen}
\psi_{\kappa}(r) = \left\{\begin{array}{ll}
\sin{\kappa r}\ , & 0 \le r \le {\pi\over\kappa}\\
0 \ , & r > {\pi\over\kappa}
\end{array}\right. \; .
\eea
We employ the $s$-wave oscillator functions
\bea
\varphi_{2n+1}(r) &=&(-1)^n \sqrt{2 n!\over\Gamma(n+3/2)} r L^{1\over 2}_n\left(r^2\right) 
e^{-{r^2\over 2}} \nonumber\\
&=&\left(\pi^{1\over 2} 2^{2n} (2n+1)!\right)^{-1/2}H_{2n+1}(r)e^{-{r^2\over 2}} \; , \nonumber
\eea
with energy $E=(2n +3/2)\hbar\Omega$. Here, $L_n^{1/2}$ denotes the
Laguerre polynomial, and it is convenient to rewrite this function in
terms of the Hermite polynomial $H_n$. We expand the
eigenfunction~(\ref{eigen}) as
\beq
\label{expand}
\psi_{\kappa}(r) = \sum_{n=0}^\infty c_{2n+1}(\kappa)\varphi_{2n+1}(r) \; .
\eeq
Before we turn to the computation of the expansion coefficients
$c_{2n+1}(\kappa)$, we consider the eigenvalue problem for the
operator $p^2$.  We have
\beq
p^2 = a^\dagger a +{1\over 2} -{1\over 2}\left(a^2 +\left(a^\dagger\right)^2\right) \;,
\eeq
where $a$ and $a^\dagger$ denote the annihilation and creation
operator for the one-dimensional harmonic oscillator, respectively.
The matrix of $p^2$ is tridiagonal in the oscillator basis.
For the matrix representation, we order the basis states as
$(\varphi_1, \varphi_3, \varphi_5, \ldots)$. Thus, the eigenvalue
problem $p^2-\kappa^2=0$ becomes a set of rows of coupled linear
equations. In an infinite basis, the eigenvector $(c_1(\kappa),
c_3(\kappa), c_5(\kappa), \ldots )$ identically satisfies every row of
the eigenvalue problem for any value of $\kappa$. In a finite basis
$(\varphi_1, \varphi_3, \varphi_5,\ldots \varphi_{2n+1})$, however,
the last row of the eigenvalue problem
%
%
\beq
\label{quant}
\left(2n+3/2 -\kappa^2\right) c_{2n+1}(\kappa) =
 {1\over
  2}\sqrt{2n}\sqrt{2 n+1} \, c_{2n-1}(\kappa) \; , 
\eeq 
can only be fulfilled for certain values of $\kappa$, and this is the
quantization condition. To solve this eigenvalue problem we need
expressions for the expansion coefficients $c_{2n+1}(\kappa)$
for $n\gg 1$. Those can be derived analytically as follows.

We rewrite the eigenfunction~(\ref{eigen}) as a Fourier transform
\beq
\psi_{\kappa} (r) = \sqrt{2\over \pi} \int\limits_0^\infty dk 
\tilde{\psi}_{\kappa}(k) \sin kr \; , 
\eeq
and expand the sine function in terms of oscillator functions as 
\beq
\sin kr = \sqrt{\pi\over 2} \sum_{n=0}^\infty (-1)^n \varphi_{2n+1}(r)
\varphi_{2n+1}(k) \; . 
\eeq
Thus, the expansion coefficients in Eq.~(\ref{expand}) are given in
terms of the Fourier transform $\tilde{\psi}_\kappa(k)$ as
\beq
\label{integ}
c_{2n+1}(\kappa) = (-1)^n\int\limits_0^\infty dk\, \tilde{\psi}_{\kappa}(k)
\varphi_{2n+1}(k) \; .
\eeq
So far, all manipulations have been exact.  We need an expression for
$c_{2n+1}(\kappa)$ for $n\gg 1$ and use the asymptotic expansion
\beq
\label{approxwf}
\varphi_{2n+1}(k) \approx {(-1)^n\sqrt{2}\over \pi^{1/4}} {(2n-1)!!\over \sqrt{(2n)!}}
\sin (\sqrt{4n+3}k) \; ,
\eeq
which is valid for $|k|\ll \sqrt{2n}$, see~\cite{gradshteyn}.
Using this approximation, one finds (making use of Fourier transforms)
\bea
\label{phi}
c_{2n+1}(\kappa) &\approx& \pi^{1/4} {(2n-1)!!\over \sqrt{(2n)!}} 
\psi_{\kappa}(\sqrt{4n+3}) \nonumber\\
&=&\pi^{1/4}{(2n-1)!!\over \sqrt{(2n)!}}
\sin (\sqrt{4n+3}\kappa) \; , 
\eea
with $\kappa \le \pi/\sqrt{4n+3}$ due to Eq.~(\ref{eigen}).

Let us return to the solution of the quantization
condition~(\ref{quant}).  We make the ansatz
\beq
\kappa = {\pi\over\sqrt{4n+3+2\Delta}} \; , 
\eeq
and must assume that $\Delta > 0$.  This ansatz is well motivated, since the
naive semiclassical estimate $\kappa = \pi/L_0$ yields $\Delta=0$. We
insert the expansion coefficients~(\ref{phi}) into the quantization
condition~(\ref{quant}) and consider its leading-order approximation
for $n\gg 1$ and $n\gg \Delta$. This yields 
\beq
\Delta = 2 
\eeq 
as the solution. Recalling that a truncation of the basis at
$\varphi_{2n+1}$ corresponds to the maximum energy
$E=(2n+3/2)\hbar\Omega$, we see that we must identify $N=2n$. 
Thus, $\kappa_{\rm min} = \pi/L_2$ is the lowest momentum (or minimum step of momentum) in a finite oscillator basis with $n \gg 1$ basis states (and not $1/b$ as stated in Ref.~\cite{Coon:2012ab}). It is clear from its very definition that $\pi/L_2$ is also (a very precise approximation of) the natural infrared cutoff in a finite oscillator basis, and that $L_2$ (and not $b$ as stated in Refs.~\cite{Stetcu:2006ey,Stetcu2010jpg}) is the radial extent of the oscillator basis and the analogue to the extent of the lattice in lattice computations~\cite{Luscher:1985dn}.”

The derivation of our key result $\kappa_{\rm min}=\pi/L_2$ is based
on the assumption that the number of shells $N$ fulfills $N\gg 1$.
Table~\ref{tab1} shows a comparison of numerical results for
$\kappa_{\rm min}$ in different model spaces. We see that
$\pi/L_2$ is a very good approximation already for $N=2$, with a
deviation of about 1\%.

\begin{table}[ht]
\begin{tabular}{|c|c|c|c|}\hline
$N$ & $\kappa_{\rm min}$ & $\pi/L_2$ & $\pi/L_0$ \\\hline
   0 & 1.2247 & 1.1874 & 1.8138\\
   2 & 0.9586 & 0.9472 & 1.1874\\
   4 & 0.8163 & 0.8112 & 0.9472\\
   6 & 0.7236 & 0.7207 & 0.8112\\
   8 & 0.6568 & 0.6551 & 0.7207\\
  10 & 0.6058 & 0.6046 & 0.6551\\
  12 & 0.5651 & 0.5642 & 0.6046\\
  14 & 0.5316 & 0.5310 & 0.5642\\
  16 & 0.5035 & 0.5031 & 0.5310\\
  18 & 0.4795 & 0.4791 & 0.5031\\
  20 & 0.4585 & 0.4582 & 0.4791\\\hline
\end{tabular}
\caption{Comparison between the lowest momentum $\kappa_{\rm min}$, $\pi/L_2$, and $\pi/L_0$ for model spaces with up to $N$ oscillator quanta.} 
\label{tab1}
\end{table}

Note that this approach can be generalized to other localized
bases. As the number of basis states is increased, the (numerical)
computation of the lowest eigenvalue of the momentum operator $p^2$
yields the box size $L$ corresponding to the employed Hilbert space,
and results can then be extrapolated according to
Eq.~(\ref{eq:IR_scaling1}).

\subsection{Scattering phase shifts}

The argument for computing scattering phase shifts is as
follows: The oscillator basis appears as a spherical box of size
$L$. For low momenta we have $L=L_2$, but at higher momentum $L$
deviates slightly from $L_2$, and can be determined from the
eigenvalues of the operator $p^2$.  Thus, the positive-energy states
computed in the oscillator basis can be used to extract phase shifts.

In a fixed harmonic oscillator basis ($N,\hbar\Omega$), the
computation of the phase shifts for a given partial wave $^{2S+1}l_J$
with orbital angular momentum $l$ proceeds as follows: First, one
computes the discrete eigenvalues $p_i^2$ of the operator $p^2$ for
orbital angular momentum $l$. Second, we need to determine the
momentum dependent box size $L_i=L(p_i)$. Assuming that the $i^{\rm
  th}$ momentum eigenstate is the $i^{\rm th}$ eigenstate of a
spherical box, we must determine the $i^{\rm th}$ zero of the
spherical Bessel function. Thus $j_l(p_iL_i/\hbar) = 0 $ determines
$L(p_i)$. We evaluate the smooth function $ L(p)$ for arbitrary
momentum $p$ by interpolating between the discrete momenta
$p_i$. Third, we compute the discrete positive energies $E_i =
\hbar^2k_i^2/(2m)$ of the neutron-proton system in relative
coordinates for the partial wave $^{2S+1}l_J$, and compute the phase
shifts from the Dirichlet boundary condition at $r=L$, i.e.
\beq
\tan\delta_l(k_i) = { j_l(k_iL(\hbar k_i)) \over \eta_l(k_iL(\hbar k_i)) 
} \; .  
\eeq
Here $\eta_l$ is the spherical Neumann function.  In practice one
repeats this procedure for several values of $\hbar\Omega$ in order to
get sufficiently many datapoints that fall onto a smooth curve.

\begin{figure}[tbh-]
\includegraphics[width=0.90\columnwidth]{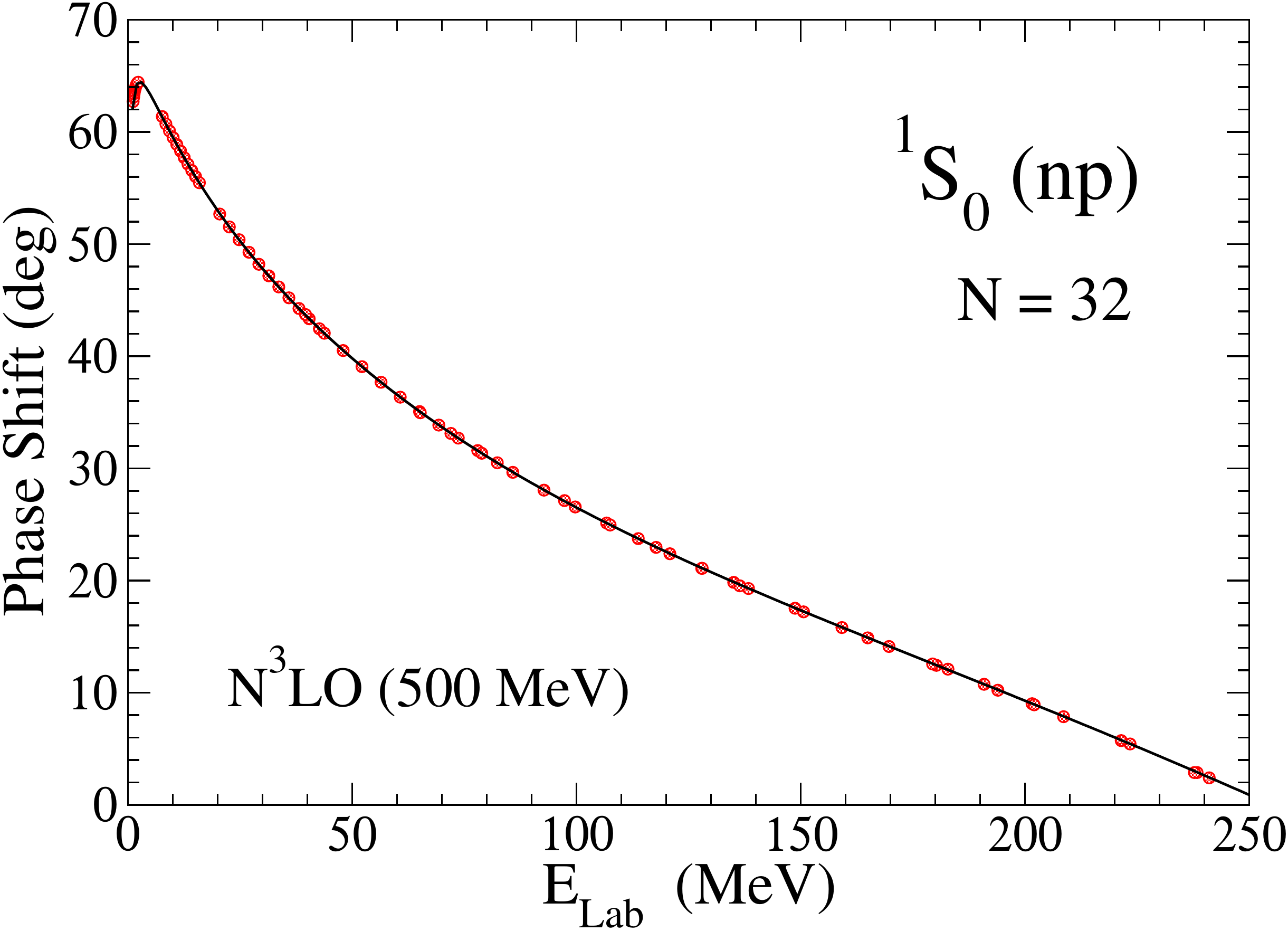}
\caption{(color online) The $^1$S$_0$ phase shifts (in degrees) of the N$^3$LO 
  chiral interaction (solid line) compared to the phase shifts
  computed directly in the harmonic oscillator basis (circles).}
\label{fig:1s0phase}
\end{figure}

\begin{figure}[tbh-]
\includegraphics[width=0.90\columnwidth]{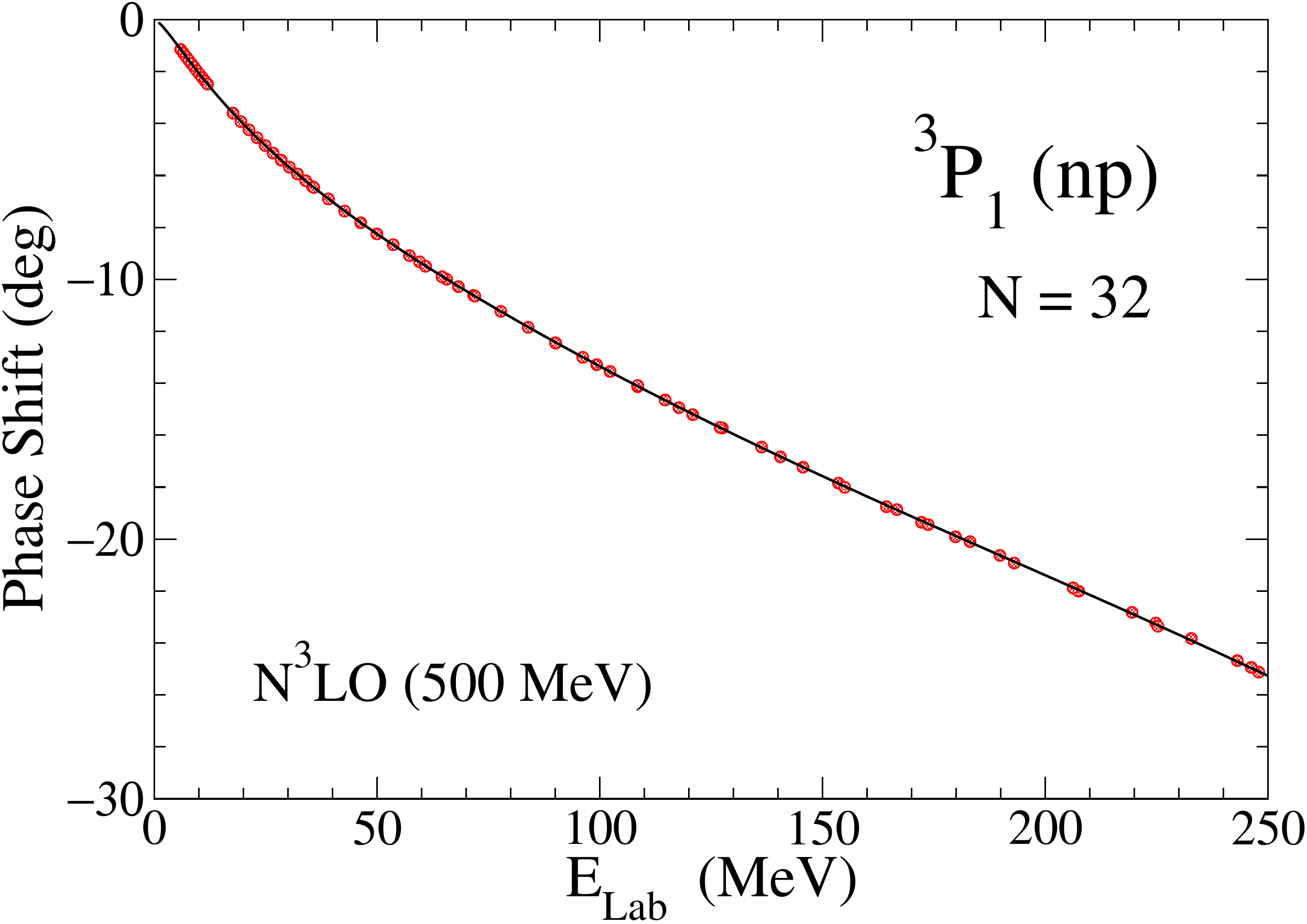}
\caption{(color online) The $^3$P$_1$ phase shifts (in degrees) of the N$^3$LO 
  chiral interaction (solid line) compared to the phase shifts
  computed directly in the harmonic oscillator basis (circles).}
\label{fig:3p1phase}
\end{figure}

As examples we compute the scattering phase shifts for the $^1$S$_0$ and $^3$P$_1$
partial waves in model spaces with $N=32$ and
$\hbar\Omega=20,22,\ldots,40$~MeV. Our calculations are based on the
Entem-Machleidt 500\,MeV chiral EFT N$^3$LO
potential~\cite{Entem:2003ft}. Figures~\ref{fig:1s0phase} and \ref{fig:3p1phase}
show the results and compares them to the numerically exact phase
shifts.  For smaller $N$ than our current choice, the
computed phase shifts start to deviate from exact phase shifts at
higher energies.  However, if one is interested only in low-energy
phase shifts and observables such as the scattering length and the
effective range, a smaller harmonic oscillator basis is sufficient.

There are other methods to compute scattering phase shifts in the
harmonic oscillator basis. Bang \emph{et al.} \cite{bang2000} used the
method of harmonic oscillator representation of scattering equations
(HORSE) for this purpose, and more recent
works~\cite{Luu2010,Stetcu2010jpg} computed phase shifts to develop an
EFT for nuclear interactions directly in the oscillator
basis~\cite{Stetcu:2006ey}. References~\cite{Luu2010,Stetcu2010jpg}
build on the results by Busch {\it et al.}~\cite{busch1998} and their
generalization~\cite{Bhattacharyya2006} to finite range corrections,
and extract scattering information from the energy shifts of bound
states in a harmonic oscillator potential. The resulting EFTs are
quite efficient for contact interactions and systems such as ultracold
trapped fermions, but nuclear potentials with a finite range require
an extrapolation of $\Omega\to 0$~\cite{Luu2010}. The approach
presented in this Subsection is more direct, as no external oscillator
potential is employed. We note that the analysis presented in this
Subsection can easily be extended to coupled channels as well.

Finally, we note again that the approach of this Subsection can be
utilized in other localized basis sets. All that is required is the
diagonalization of the operator $p^2$ in the employed basis set, which
yields the (momentum dependent) box size.

\subsection{Functional dependence of extrapolation}
\label{subsec:extrapolation_form}
  
The extrapolation formula~(\ref{eq:IR_scaling1}) with $L=L_2$ is
theoretically founded. How well can the specific form of this
extrapolation be distinguished from other popular empirical choices?
To address this question, we test possible \emph{functional}
dependences of the energy correction $\Delta E$ on $L$.  The most
common extrapolation schemes employ an exponential in $N$ (or
equivalently a Gaussian dependence on $L$),
\beq
    E(N) = \Einf + C_N e^{-b_N N}
    \;,
    \label{eq:gauss}
\eeq
where $C_N$ and $b_N$ are determined separately for each $\hw$ (with
the option of a constrained fit of a common $\Einf$ for special \hw\
values).  Thus, unlike the extrapolation based on $L$, there is no
universal variable and no distinction between IR and UV regions in
\hw.  However, empirically the form in Eq.~\eqref{eq:gauss} seems to
work quite
well~\cite{Hagen:2007hi,Bogner:2007rx,Forssen:2008qp,Maris:2008ax,Roth:2009cw}.
Recently, Tolle {\it et al.}~\cite{Tolle:2012cx} investigated the
convergence properties of genuine and smeared contact interactions in
an effective theory of trapped bosons and found that the smearing
changed a power law dependence of the convergence to an exponential
dependence.  Here we will consider all three functional dependences on
$L$: exponential, Gaussian, and power law.

\begin{figure}[tbh-!]
\includegraphics[width=0.95\columnwidth]{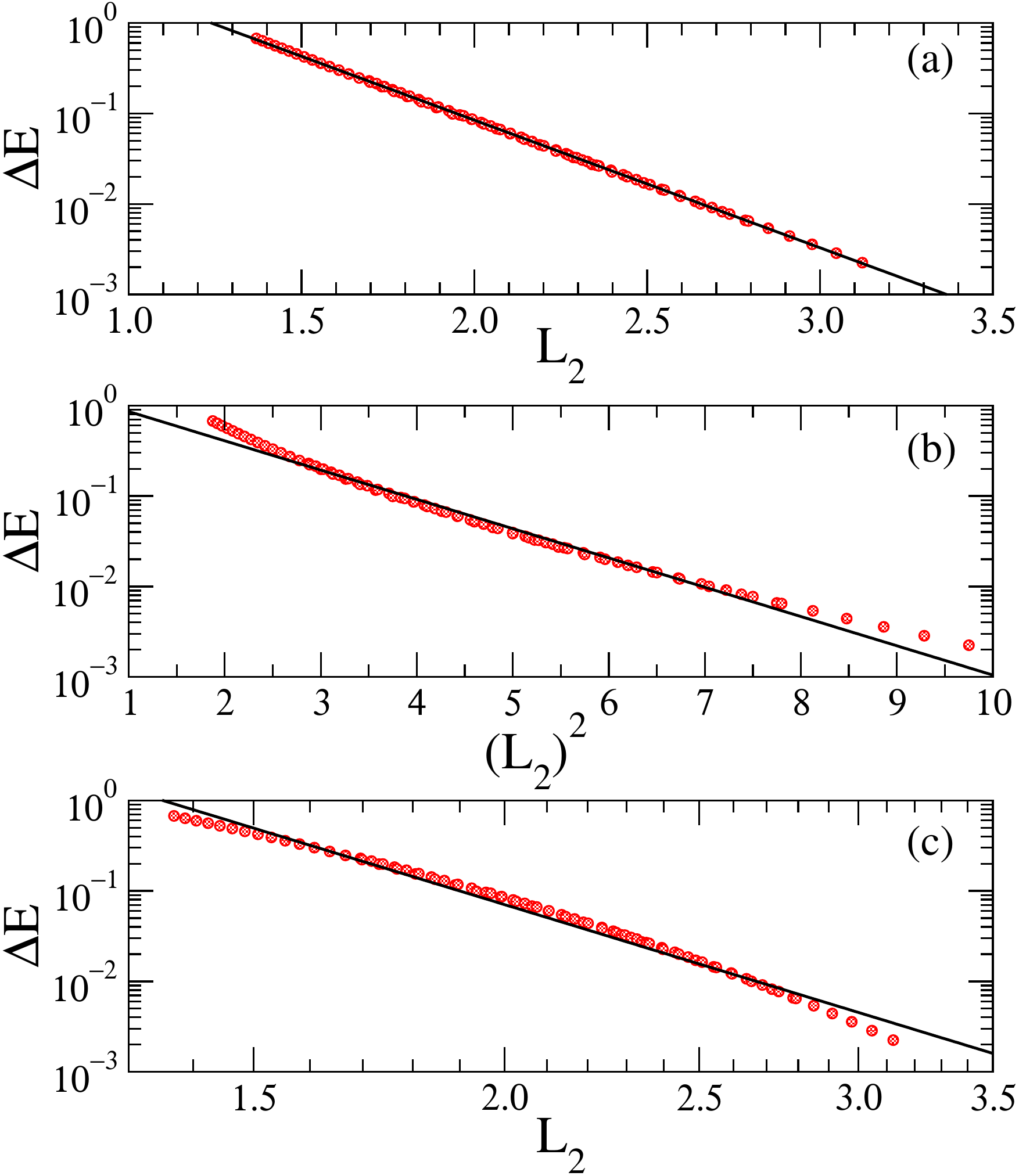}
\caption{(color online) The IR energy correction $\Delta E_L$ versus $L_2$
for a Gaussian potential well Eq.~\eqref{eq:Vg} with
$V_0 = 5$ (and $\hbar = \mu = R=1$) using a wide range of $N$ and $\hw$.
The energies are fitted with (a) exponential, (b) Gaussian,
and (c) power law dependence on $L_2$.}
\label{fig:Vgauss_other_forms}
\end{figure}

\begin{figure}[tb!]
\includegraphics[width=0.95\columnwidth]{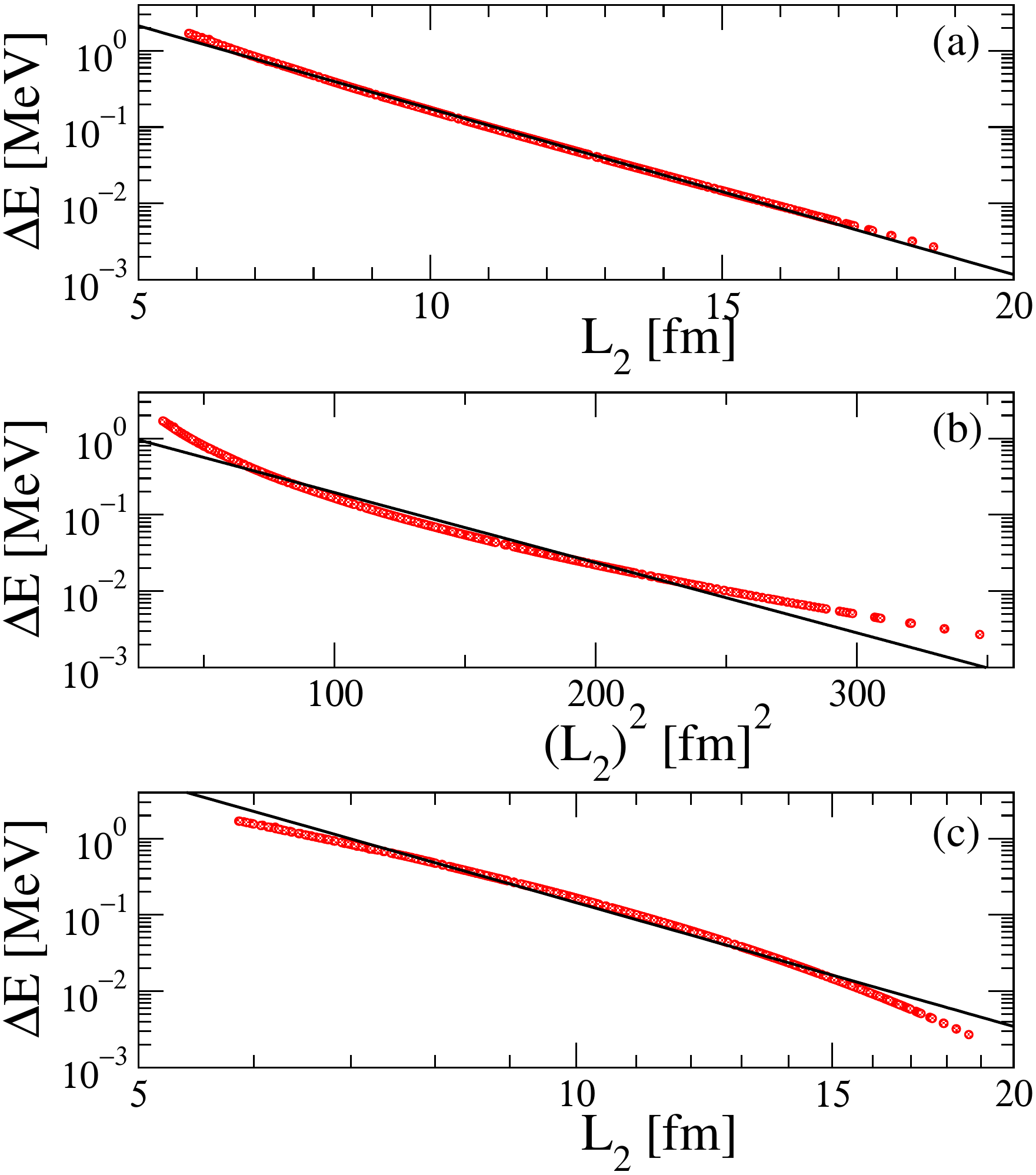}
\caption{(color online) The IR energy correction $\Delta E_L$
 versus $L_2$ for the deuteron calculated with the chiral EFT potential from
Ref.~\cite{Entem:2003ft} using a wide range of $N$ and $\hw$.
The energies are fitted with (a) exponential, (b) Gaussian,
and (c) power law dependence on $L_2$.}
\label{fig:deuteron_other_forms}
\end{figure}

A purely empirical test can be made for our models and the deuteron
because we can calculate the exact $\Einf$, plot $\Delta E(L_2) \equiv
E(L_2) - \Einf$ against $L_2$, and then attempt to fit each of the
three choices of $\Delta E(L_2)$.  Figure~\ref{fig:Vgauss_other_forms}
shows the results for a representative model potential (a Gaussian)
with moderate depth while Fig.~\ref{fig:deuteron_other_forms} shows
the results for the deuteron.  The plots are made so that the
candidate form would yield a straight line if followed precisely.  We
see that the exponential form is an excellent fit for the model
throughout the range of $L_2$ and a reasonable but not perfect fit for
the deuteron. For the deuteron, the weak binding is a challenge as it
requires very large values of $L_2$ for extrapolations. Corrections to
weak binding will be derived in Section~\ref{sec:universality}.  In
contrast to the exponential extrapolation, Gaussian and power law fits fail
over the full range of $L_2$.  This is consistent with 
Tolle {\it et al.}~\cite{Tolle:2012cx}.  For limited ranges of $L_2$ a Gaussian does
provide a reasonable fit (and should give a good extrapolation for
$\Einf$ if close enough to convergence), but not globally.

At this stage we have empirically verified the usefulness of the
extrapolation~\eqref{eq:IR_scaling1} in a very controlled setting.
This corroborates the study in Ref.~\cite{Furnstahl:2012qg} and applications
in Refs.~\cite{Soma:2012zd,Hergert:2012nb}.  The fit result for
$\kinf$ has generally been quantitatively consistent with nucleon
separation energies (note, however, the case of $^6$He in
Ref.~\cite{Furnstahl:2012qg}), but the constant $A$ was not identified
with physical quantities.  The next section will express $A$ in terms
of observables for the two-particle system and present corrections 
to the extrapolation law~(\ref{eq:IR_scaling1}).


\section{Universal formulas for IR corrections}
\label{sec:universality}
   
In this Section we revisit the derivation of
Eq.~\eqref{eq:IR_scaling1} and obtain an expression for the
coefficient $A$ in terms of the bound-state asymptotic normalization
coefficient (ANC) $\ANC$ and $\kinf$.  This is in close analogy to
correction formulas for energies calculated with lattice
regularization for periodic and hard wall boundary
conditions~\cite{Luscher:1985dn,Lee:2010km,Konig:2011ti,Pine:2012zv}.
Because $\kinf$ and $\ANC$ are measurable, the result is universal in
the sense that it is the same for any potential that reproduces the
experimental observables for the bound state.  The parameters in
Eq.~\eqref{eq:IR_scaling1} can be fully predicted and tested against
precise numerical fits for both our models and the deuteron, which is
carried out in Section~\ref{sec:tests}.  Corrections to
Eq.~\eqref{eq:IR_scaling1} derived below are found to be
quantitatively important for shallow bound states.

\subsection{Linear energy approximation}

Our first approximation to the IR correction
is based on what is known in quantum chemistry
as the linear energy method~\cite{Djajaputra:2000aa}.  Given a
hard-wall boundary condition at $r=L$ beyond the range of the
potential, we write the energy compared to that for $L=\infty$ as
\beq
 E_L = E_{\infty}+\Delta E_L
 \;.
\eeq
We seek an estimate for $\Delta E_L$, which is assumed to be small,
based on an expansion of the wave function in $\Delta E_L$.  Let
$u_E(r)$ be a radial solution with regular boundary condition at the
origin and energy $E$.  For convenience in using standard quantum
scattering formalism below, we choose the normalization corresponding
to what is called the ``regular solution'' in
Ref.~\cite{taylor2006scattering}, which means that $u_E(0) = 0$ and
the slope at the origin is unity for all $E$.  We denote the
particular solutions $u_{E_L}(r)\equiv u_L(r)$ and $u_{\Einf}(r)
\equiv u_\infty (r)$. Then there is a smooth expansion of $u_E$ about
$E=\Einf$ at fixed $r$, so we approximate~\cite{Djajaputra:2000aa}
\beq
  u_L(r)\approx u_\infty (r) + \Delta E_L 
  \left.\frac{du_E(r)}{dE}\right|_{E_{\infty}} 
  + \mathcal{O}(\Delta E_L^2) \;,
  \label{eq:linear_energy_approx}
\eeq
for $r\leq L$. 
By evaluating Eq.~\eqref{eq:linear_energy_approx} at $r=L$ with the
boundary condition $u_L(L)=0$, we find
\beq
  \Delta E_L \approx -u_\infty(L) \left(\left.\frac{d u_E(L)}
  {dE}\right|_{\Einf}\right)^{-1}
  \;,
  \label{eq:Delta_EL}
\eeq
which is the estimate for the IR correction.

\begin{figure}[b!]
\includegraphics[width=0.95\columnwidth]{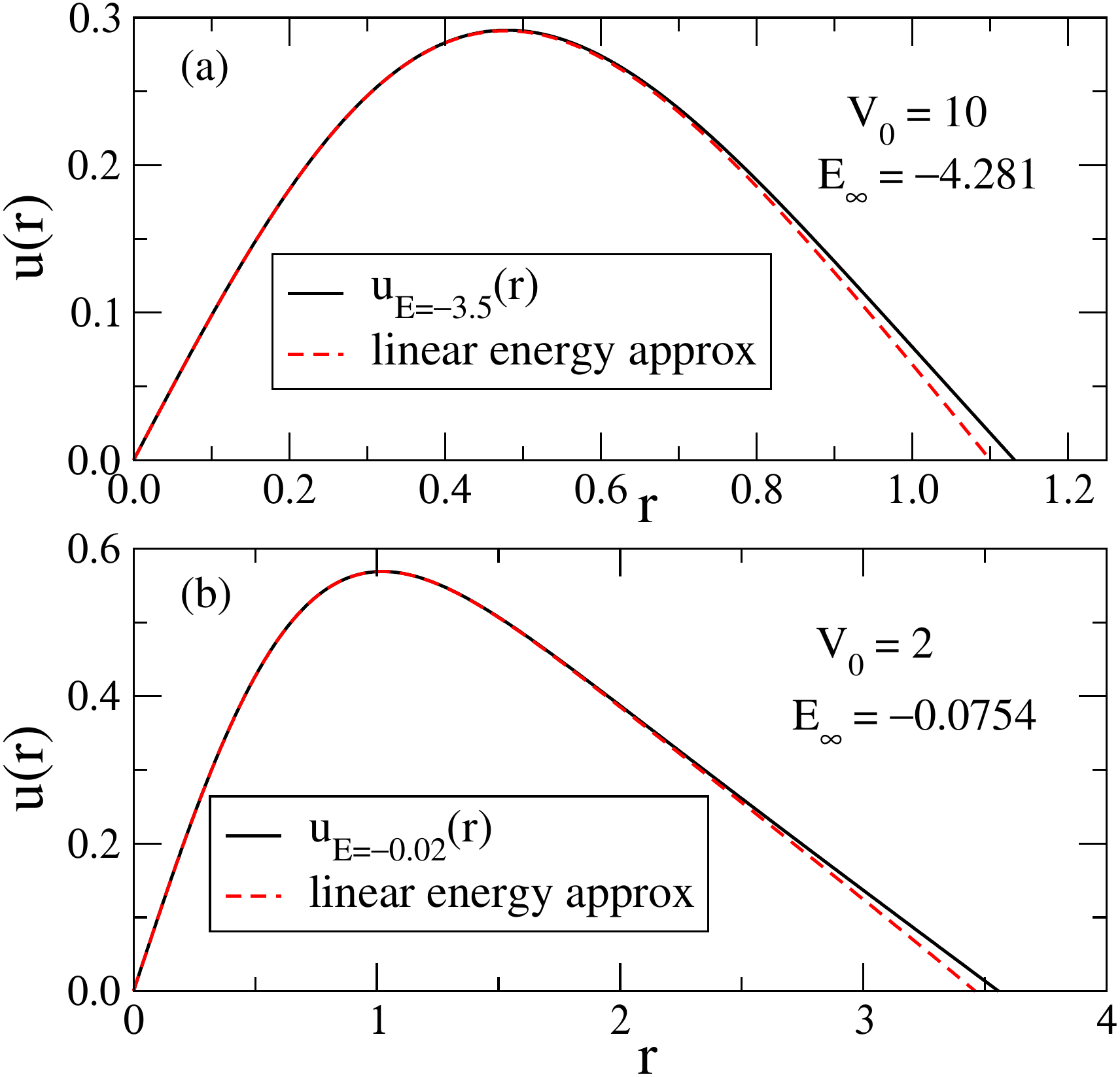}
\caption{(color online) Testing the linear energy approximation
  Eq.~\eqref{eq:linear_energy_approx} for (a) deep ($V_0=10$) and (b)
  shallow ($V_0=2$) Gaussian potential well Eq.~\eqref{eq:Vg} ($\hbar = \mu = R=1$).
  The solid lines are the exact solutions $u_L(r)$ for energies $-3.5$
  and $-0.020$, respectively, whose zero crossings determine the
  corresponding values for $L$.}
\label{fig:linear_energy_approx}
\end{figure}

We can check the accuracy of the linear energy
approximation~(\ref{eq:linear_energy_approx}) by numerically solving
the Schr\"odinger equation with a specified energy.  This determines
$L$ as the radius at which the resulting wave function vanishes. Then
we compare this wave function for $r \leq L$ to the right side of
Eq.~(\ref{eq:linear_energy_approx}), with the derivative calculated
numerically.  Figure~\ref{fig:linear_energy_approx} shows
representative examples for a deep and shallow Gaussian potential.  In
these examples and other cases, the approximation to the wave function
is good, particularly in the interior.  The estimates for $\Delta E_L$
using the right side of Eq.~\eqref{eq:Delta_EL} are within a few to
ten percent: 0.68 versus 0.70 and 0.050 versus 0.055 for the two
cases.

The good approximation to the wave function suggests that for the
calculation of other observables the linear energy approximation will
be useful.  For observables most sensitive to the long distance
(outer) part of the wave function, such as the radius, this has
already been shown to be true~\cite{Furnstahl:2012qg}. But the good
approximation to the wave function at small $r$ means that corrections
for short-range observables should also be controlled, with the
dominant contribution in an extrapolation formula coming from the
normalization.

\subsection{Complete IR scaling} 

Next we derive an expression for the derivative in
Eq.~\eqref{eq:Delta_EL}.  We assume we have a single partial-wave
channel and reserve the generalization to coupled channels 
(e.g., for a complete treatment of the deuteron) for future
work.  For general $E < 0$, the asymptotic form of the radial
wave function for $r$ greater than the range of the potential is
(using the notation of Ref.~\cite{Furnstahl:2012qg})
\beq
  u_E(r)\overset{r \gg R}\longrightarrow A_E(e^{-k_E r}+\alpha_E e^{+k_E r})
  \;,
  \label{eq:asymptotic_form_uE}
\eeq
with $u_\infty(r)\overset{r \gg R}\longrightarrow \Ainf e^{-\kinf r}$
for $E=\Einf$. We take the derivative of
Eq.~\eqref{eq:asymptotic_form_uE} with respect to energy, evaluate at
$E=\Einf$ using $\alpha_{\Einf}=0$ and $dk_E/dE = -\mu/(\hbar^2 k_E)$,
to find
\bea
    \left.\frac{d u_E(r)}{d E}\right|_{\Einf} &=&
    \Ainf \left.\frac{d \alpha_E}{d E}\right|_{\Einf} e^{+\kinf r} 
    +
    \Ainf \frac{\mu}{\hbar^2}\frac{r}{\kinf} e^{-\kinf r}
    \nonumber \\  & & \null +
    \left.\frac{d A_E}{d E}\right|_{\Einf} e^{-\kinf r} 
    \;.
  \label{eq:entire_correction}
\eea
We now evaluate at $r=L$ and anticipate that the $e^{+\kinf L}$ term
dominates:
\beq
  \left.\frac{d u_E(L)}{d E}\right|_{\Einf} 
  \approx \Ainf \left.\frac{d\alpha_E}{d E}\right|_{\Einf} e^{+\kinf L}   
   + \mathcal{O}(e^{-\kinf L})
   \;.
  \label{eq:substitute_this}
\eeq
Substituting Eq.~\eqref{eq:substitute_this} into
Eq.~\eqref{eq:Delta_EL}, we obtain
\beq \Delta E_L \approx -\left[ \left.\frac{d \alpha_E}{d
      E}\right|_{\Einf}\right]^{-1} e^{-2 \kinf L} + \mathcal{O}(e^{-4
  \kinf L}) \;, \eeq which is in the form of
Eq.~\eqref{eq:IR_scaling1}.  Note that this result is independent of
the normalization of the wave function.

To calculate the derivative explicitly, we turn to scattering theory,
following the notation and discussion in
Ref.~\cite{taylor2006scattering}.  In particular, the asymptotic form
of the regular scattering wave function $\phi_{l,k}$ for orbital
angular momentum $l$ and for positive energy $E \equiv \hbar^2 k^2/2\mu$
is given in terms of the Jost function
$\Jost_l(k)$~\cite{taylor2006scattering},
\beq
  \phi_{l,k}(r) \longrightarrow 
    \frac{i}{2}[\Jost_l(k)\hat{h}_l ^{-}(k r)-\Jost_l(-k)\hat{h}_l^{+}(k r)]
    \;,
    \label{eq:phi_asymp}
\eeq
where the $\hat{h}_l^{\pm}$ functions (related to Hankel functions)
behave asymptotically as
\beq
\hat{h}_l ^{\pm}(k r)
    \overset{r\rightarrow\infty}\longrightarrow e^{\pm i (k r - l \pi/2)}
    \;.
\eeq
The ratio of the Jost functions appearing in Eq.~\eqref{eq:phi_asymp}
gives the partial wave $S$-matrix $s_l(k)$:
\beq
   s_l(k) = \frac{\Jost_l(-k)}{\Jost_l(+k)}
   \;,
   \label{eq:s_l}
\eeq
which is in turn related to the partial-wave scattering amplitude
$f_l(k)$ by
\beq
  f_l(k) = \frac{s_l(k) - 1}{2i k}
  \;.
  \label{eq:f_l}
\eeq
We will restrict ourselves to $l=0$ for simplicity; the generalization
to higher $l$ is straightforward.

To apply Eq.~\eqref{eq:phi_asymp} to negative energies, we
analytically continue from real to (positive) imaginary $k$.  So,
\bea
   \phi_{0,ik_E}(r)
       &\overset{r \gg R}\longrightarrow&
         \frac{i}{2}\bigl(
	    \Jost_0(ik_E)e^{k_E r}-\Jost_0(-ik_E) e^{-k_E r} \bigr)
	 \nonumber \\
       &=& 
       -\frac{i}{2}\Jost_0(-ik_E) \bigl(
           e^{-k_E r} - \frac{\Jost_0(-ik_E)}{\Jost_0(ik_E)} e^{k_E r}
         \bigr)
       	\;, 
	\nonumber \\
  \label{eq:asymptotic_form_phiE}
\eea
where $R$ is the range of the potential.  Upon comparing to
Eq.~\eqref{eq:asymptotic_form_uE} we conclude that
\beq
  \alpha_E = -\frac{\Jost_0(ik_E)}{\Jost_0(-ik_E)}
   = -\frac{1}{s_0(ik_E)} \;.
  \label{eq:alphaE_and_S}
\eeq
Note that Eq.~\eqref{eq:alphaE_and_S} is consistent with the
bound-state limit of Eq.~\eqref{eq:asymptotic_form_uE}: at a bound
state where $\Einf = -\hbar^2 \kinf^2/2\mu$ there is a simple pole in the $S$
matrix, which means $\alpha_E = 0$ as expected (no exponentially
rising piece).

From Ref.~\cite{taylor2006scattering} we learn that the residue as a
function of $E$ of the partial wave amplitude $f_l(E)$ at the
bound-state pole is $(-1)^{l+1} \ANC^2 \hbar^2/2 \mu$, where $\ANC$ is
the ANC.  The ANC is defined by the large-$r$ behavior of the
\emph{normalized} bound-state wave function:
\beq
  u_{\rm norm}(r)\overset{r\gg R}\longrightarrow \ANC e^{-\kinf r}
  \;.
  \label{eq:definition_of_ANC}
\eeq
Thus, near the bound-state pole (with $E = \hbar^2 k^2/2\mu$),
\beq
  f_0(k)  \approx  \frac{- \hbar^2\ANC^2}{2\mu (E-\Einf)}  
      =  \frac{-\ANC^2}{k^2 + \kinf^2}
   \;.
\eeq
or, using Eqs.~\eqref{eq:f_l} and \eqref{eq:alphaE_and_S},
%
%
%
\beq  
  \alpha_E(k) \approx -\frac{k^2+\kinf^2}{k^2+\kinf^2-2 i k \ANC^2}
  \;.
  \label{eq:alphaE_k}
\eeq
Now,
\beq
  \left.\frac{d \alpha_E}{d E}
  \right|_{\Einf}=\frac{d\alpha_E/dk\vert_{k=i\kinf}}
  {dE/dk|_{k=i\kinf}}
  \;,
\eeq
so using Eq.~\eqref{eq:alphaE_k} we find
\beq
  \left.\frac{d\alpha_E}{d k}\right|_{k=i\kinf}=\frac{-i}{\ANC^2}
  \;,
\eeq
and therefore
\beq
  \left.\frac{d\alpha_E}{dE}\right|_{\Einf}=\frac{-\mu}{\hbar^2 \kinf \ANC^2}
  \;.
\eeq
Putting it all together, we have
\beq
  \Delta E_L = \frac{\hbar^2 \kinf \ANC^2}{\mu} e^{-2 \kinf L}
    + \mathcal{O}(e^{-4 \kinf L})
    \;,
    \label{eq:complete_IR_scaling}
\eeq
in agreement with Eq.~\eqref{eq:IR_scaling1}, but now we have
identified $A = \hbar^2\kinf\ANC^2/\mu$.

If we apply this correction for a weakly bound state, such that
$\kinf$ is small, we may not be justified in neglecting the second
term on the right side of Eq.~\eqref{eq:entire_correction}.  If we
keep it instead, then Eq.~\eqref{eq:substitute_this} becomes
\beq
  \left.\frac{d u_E(L)}{d E}\right|_{\Einf} 
  \approx \Ainf e^{+\kinf L}   
  \left(
  \left.\frac{d\alpha_E}{d E}\right|_{\Einf}  
    + \frac{mL}{\kinf} e^{-2\kinf L}
  \right)
  \;,
  \label{eq:improved_substitute_this}
\eeq
and we have a modified infrared scaling given by
\beq
 (\Delta E_L)_{\rm mod} = \frac{\hbar^2 \kinf \ANC ^2}{\mu} 
   \frac{e^{-2 \kinf L}}{(1-\ANC^2 L e^{-2 \kinf L})}
   \;.
  \label{eq:IR_scaling}
\eeq
We will test both Eqs.~\eqref{eq:complete_IR_scaling} 
and \eqref{eq:IR_scaling} in Section~\ref{sec:tests}.

\subsection{Relation to L\"uscher-type formulas}

Starting with the seminal work of L\"uscher~\cite{Luscher:1985dn}, a
wide variety of formulas have been derived for the energy shift of
bound states in finite-volume lattice calculations.  The usual
application is to simulations that use periodic boundary conditions in
cubic boxes (e.g., see Ref.~\cite{Konig:2011ti}).  The recent work by
Pine and Lee~\cite{Lee:2010km,Pine:2012zv} extend the derivation to
hard-wall boundary conditions using effective field theory for
zero-range interactions and the method of images.  The result for
$\Delta E_L$ in a three-dimensional cubic box has a different
functional form than found here (the leading exponential is multiplied
by $1/L$ with that geometry) and the subleading corrections are
parametrically larger.

However, because the HO truncation we consider is in partial waves,
the one-dimensional analysis and formula from Ref.~\cite{Pine:2012zv}
are applicable (because $\kinf$ and $\ANC$ are asymptotic quantities,
the result for zero-range interaction is actually general for
short-range interactions).  The method of images can be applied in a
one-dimensional box of size $2L$ after specializing to a particular
partial wave and then extending the space to odd solutions in $r$ from
$-\infty$ to $+\infty$.  The leading-order finite-volume correction
agrees with Eq.~\eqref{eq:complete_IR_scaling}, and the first omitted term
is of the same order.
The methods presented in \cite{Lee:2010km,Pine:2012zv}
can be used to extend the present formulas to higher
orders and more general cases, including coupled channels.


\section{Tests of IR correction formulas}
  \label{sec:tests}

In this Section we test direct fits of Eq.~\eqref{eq:IR_scaling1},
which has three parameters, and the specialized expressions for
$\Delta E_L$ in Eqs.~\eqref{eq:complete_IR_scaling} and
\eqref{eq:IR_scaling}, which have no free parameters if we take
$\kinf$ and $\ANC$ from the exact solutions.  Based on the results
presented in Sect.~\ref{sec:HO_vs_Dirichlet}, we use $L_2$ in all our
further analyses.
It is important that we isolate the IR corrections in making these
tests.  The truncation in the HO basis also introduces an
ultraviolet error inversely proportional to the ultraviolet cutoff
$\Lambda_{\rm UV} \approx \sqrt{2 \mu \hbar \Omega (N+3/2)}$.  In the
results here we use combinations of $\hbar \Omega$ and $N$ values such
that the UV error in each case can be neglected compared to the IR
error. (This is verified quantitatively by using a fit ansatz from
Ref.~\cite{Furnstahl:2012qg} for the UV correction, which is assumed
to be independent of the IR correction.)

For each of the model potentials, the radial Schr\"odinger equation is
accurately solved numerically in coordinate space for the energy,
which yields $\kinf$, and the wave functions.  The asymptotic
normalization coefficient $\ANC$ is found by multiplying the wave
function by $e^{\kinf r}$ and reading off its asymptotic value.  This
is illustrated in the inset of Fig.~\ref{fig:IR_quartic_inset_ANC},
which also shows the onset of the plateau that defines the asymptotic
region in $L_2$ where we expect our correction formulas to hold.  For
the deuteron, the Hamiltonian is diagonalized in
momentum space to find $\kinf$, and then an extrapolation to the pole
is used to find the $s$-wave and $d$-wave ANCs~\cite{Amado:1979zz}.
In the present work we use only the $s$-wave ANC for the deuteron.
  
\subsection{Universal properties}     
\label{subsec:universal}
    
The derivations in Section~\ref{sec:universality} imply that the
energy corrections should have the same exponential form and
functional dependence on the radius $L$ at which the wave
function is zero, independent of the potential and for
any bound state (although the relationship between $L$ and
the oscillator determined $L_2$ is energy dependent).
However, there are corrections to
Eq.~\eqref{eq:complete_IR_scaling} that become increasingly important
if $L$ is not sufficiently large.  Equation~\eqref{eq:IR_scaling}
incorporates one such correction but we also have beyond-linear energy
corrections and the third term in Eq.~\eqref{eq:entire_correction}.
Here we make some representative tests of a direct fit of
Eq.~\eqref{eq:IR_scaling1} in comparison to applying
Eqs.~\eqref{eq:complete_IR_scaling} and \eqref{eq:IR_scaling}.

\begin{figure}
  \includegraphics[width=0.95\columnwidth]{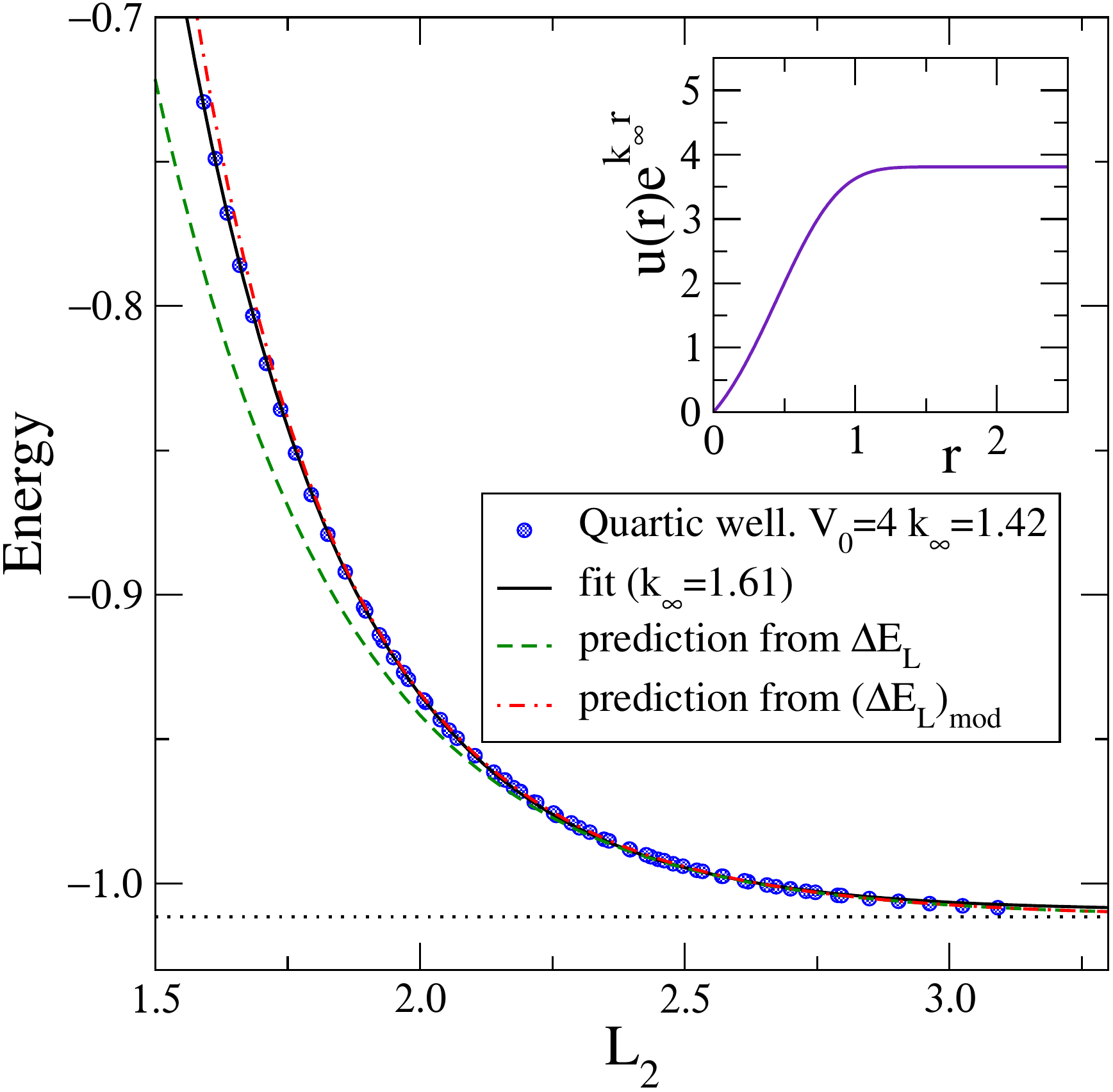}
  \caption{(color online) Energy versus $L_2$ for a quartic potential
    well Eq.~\eqref{eq:Vq} for a wide range of $N$ and $\hw$
    (circles) ($\hbar = \mu = R=1$).  The solid line is a fit to Eq.~\eqref{eq:IR_scaling1}
    with $A$, $\kinf$ and $\Einf$ as fit parameters while the dashed and
    dot-dashed lines are predictions from
    Eqs.~\eqref{eq:complete_IR_scaling} and Eq.~\eqref{eq:IR_scaling}.
    The horizontal line is the exact energy, $\Einf=-1.0115$.
    The inset illustrates the calculation of the asymptotic
    normalization coefficient (ANC) from the (normalized) wave
    function.}
\label{fig:IR_quartic_inset_ANC}
\end{figure}

\begin{figure}
\includegraphics[width=0.95\columnwidth]{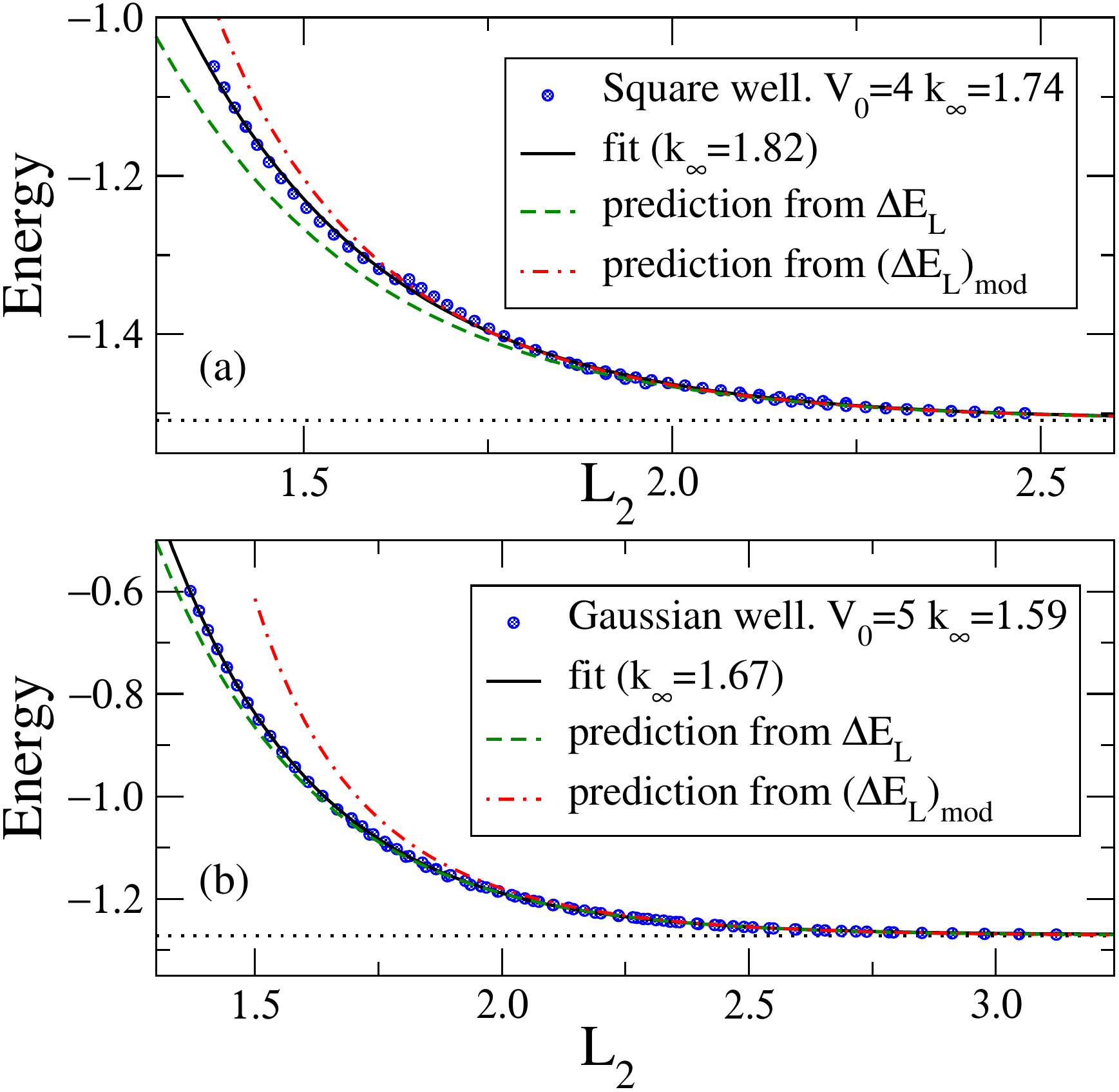}
\caption{(color online) Energy versus $L_2$ for moderate-depth (a) square
  well Eq.~\eqref{eq:Vsw} and for (b) Gaussian potential
  well Eq.~\eqref{eq:Vg} ($\hbar = \mu = R=1$) for a wide range of $N$ and $\hw$ (circles).  The solid line
  is a fit to Eq.~\eqref{eq:IR_scaling1} with $A$, $\kinf$ and $\Einf$ as
  fit parameters while the dashed and dot-dashed lines are
  predictions from Eqs.~\eqref{eq:complete_IR_scaling} and
  Eq.~\eqref{eq:IR_scaling}. The horizontal dotted lines are the exact energies;
   square well: $\Einf = -1.5088$, Gaussian well: $\Einf = -1.2717$}
\label{fig:universal_sq_gauss_wells}
\end{figure}

Figure~\ref{fig:IR_quartic_inset_ANC} shows results for a quartic
potential with a moderate depth.  The fit to
Eq.~\eqref{eq:IR_scaling1} is very good over a large range in $L_2$
for which the energy changes by 30\%, and the prediction for $\Einf$
is accurate to 0.2\%.  However, the fit value of $\kinf$ is 1.61
compared to the exact value of 1.42.  The dashed curve shows the
prediction from Eq.~\eqref{eq:complete_IR_scaling} using the exact
$\kinf$ and $\ANC$.  It is evident that the approximation is very good
above $L_2 > 2$ but increasingly deviates at smaller $L_2$.  The
modified energy correction from Eq.~\eqref{eq:IR_scaling} (dot-dashed
curve) matches the energy results at the same level as the fit.

In Fig.~\ref{fig:universal_sq_gauss_wells}, examples are shown for
square well and Gaussian potentials with a moderate depth.  Again we
find a good fit to an exponential fall-off in $L_2$, but in these
cases not only are the energies well predicted (again to better than
0.2\%) but the fit values of $\kinf$ are within 5\% of the exact
results.  
However, the prediction from Eq.~\eqref{eq:IR_scaling} actually degrades the
agreement for the Gaussian well compared to the prediction from
Eq.~\eqref{eq:complete_IR_scaling}.  Further investigation in these
cases reveals that the contributions from the second and third terms
in Eq.~\eqref{eq:entire_correction} are of comparable size and
opposite sign.  Therefore, keeping only one of them is
counterproductive.

\begin{figure}
\includegraphics[width=0.95\columnwidth]{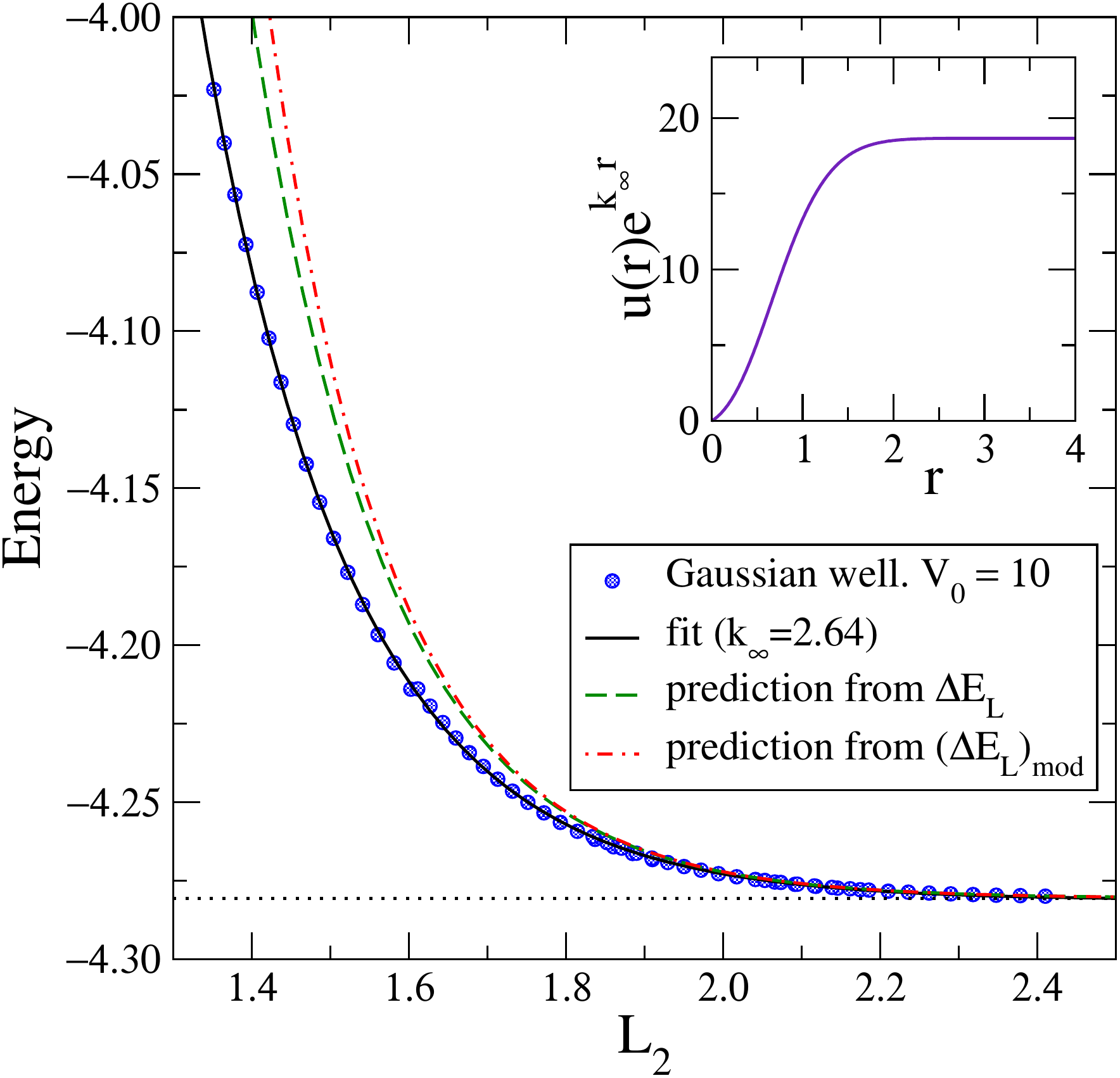}
\caption{(color online) Energy versus $L_2$ for the deeply bound
  ground state of a Gaussian potential for a wide range of $N$ and
  $\hw$ (circles) ($\hbar = \mu = R=1$).  These are compared to the predictions of
  Eq.~\eqref{eq:complete_IR_scaling} (dashed) and
  Eq.~\eqref{eq:IR_scaling} (dot-dashed). The solid line
  is a fit to Eq.~\eqref{eq:IR_scaling1} with $A$, $\kinf$ and $\Einf$ as
  fit parameters. The horizontal dotted line
  is the exact energy, $\Einf=-4.2806$.}
\label{fig:deep_gauss_ANC_inset}
\end{figure}

\begin{figure}
\includegraphics[width=0.95\columnwidth]{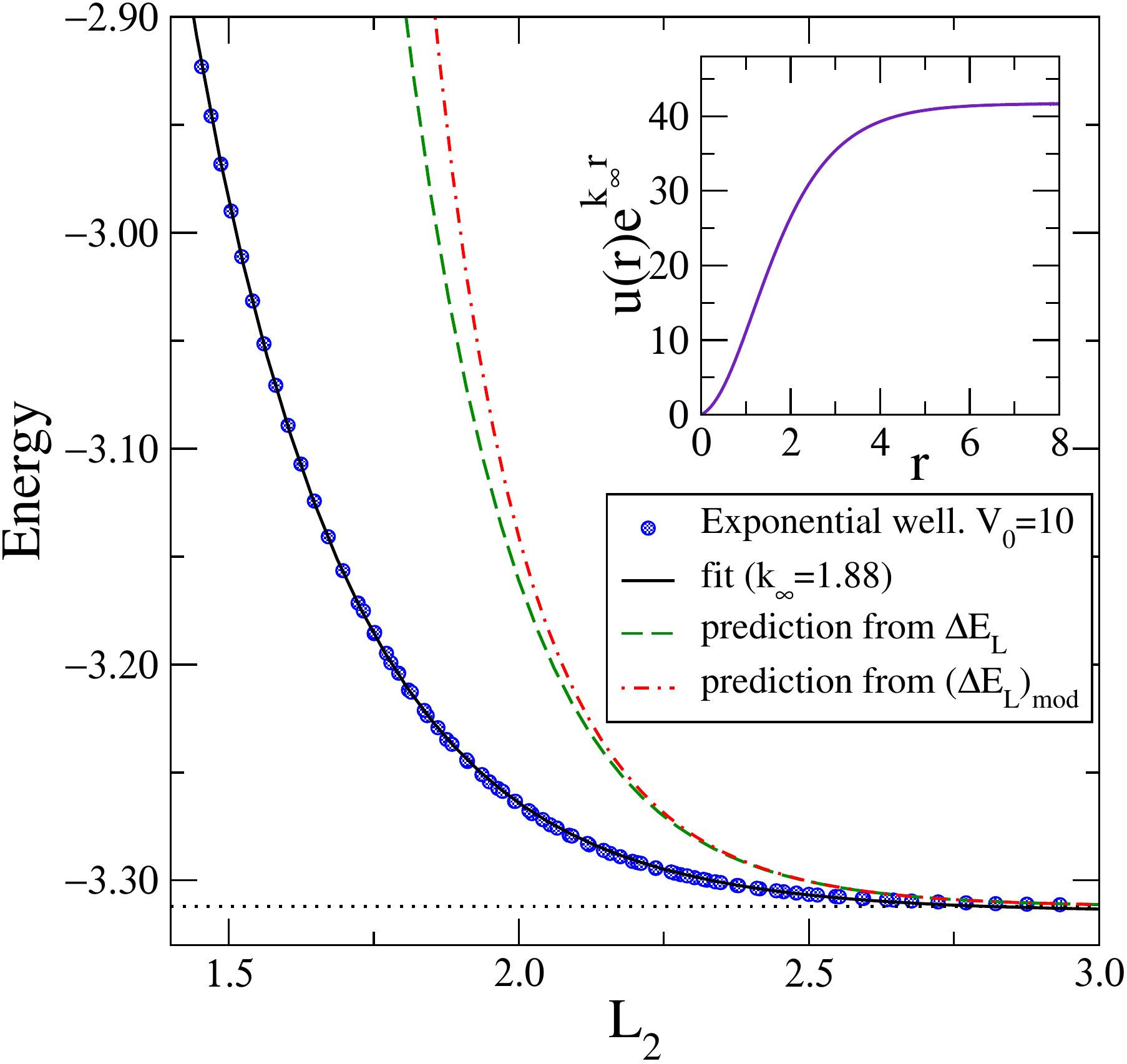}
\caption{(color online) Energy versus $L_2$ for the deeply bound
  ground state of an exponential potential well for a wide range of
  $N$ and $\hw$ (circles) ($\hbar = \mu = R=1$). These are compared to the predictions of
  Eq.~\eqref{eq:complete_IR_scaling} (dashed) and
  Eq.~\eqref{eq:IR_scaling} (dot-dashed). The solid line
  is a fit to Eq.~\eqref{eq:IR_scaling1} with $A$, $\kinf$ and $\Einf$ as
  fit parameters. The horizontal dotted line
  is the exact energy, $\Einf=-3.3121$.}\label{fig:deep_exp_ANC_inset}
\end{figure}

For deeply bound states, Eqs.~\eqref{eq:complete_IR_scaling} and
\eqref{eq:IR_scaling} can fail for a different reason.  The error in
Eq.~\eqref{eq:complete_IR_scaling} is proportional to $e^{-4 \kinf
  L}$, so one might expect that the prediction to become increasingly
accurate as the state becomes more bound. However, as seen in
Figs.~\ref{fig:deep_gauss_ANC_inset} and \ref{fig:deep_exp_ANC_inset},
results for deep Gaussian and exponential potential wells do not match
this expectation.  In deriving the energy corrections we used
the asymptotic form of the wave functions.  This is valid only in the
region $r \gg R$, where $R$ is the range of the potential.  The
potentials at the smaller values of $L_2$ shown in the figures are not
negligible.  Indeed, it is evident from the insets in
Figs.~\ref{fig:deep_gauss_ANC_inset} and \ref{fig:deep_exp_ANC_inset}
that we are not in the asymptotic region for those values of $L$.  
The lesson is that when applying the IR extrapolation schemes
discussed in the present paper we need to make sure that the two
conditions for its applicability are fulfilled. First, we need $N$
sufficiently large
for $L_2$ to be the correct box size
(see Table~\ref{tab1}). Second we need $L_2$ to be the
largest length scale in the problem under consideration. 
   
\begin{figure}
\includegraphics[width=0.95\columnwidth]{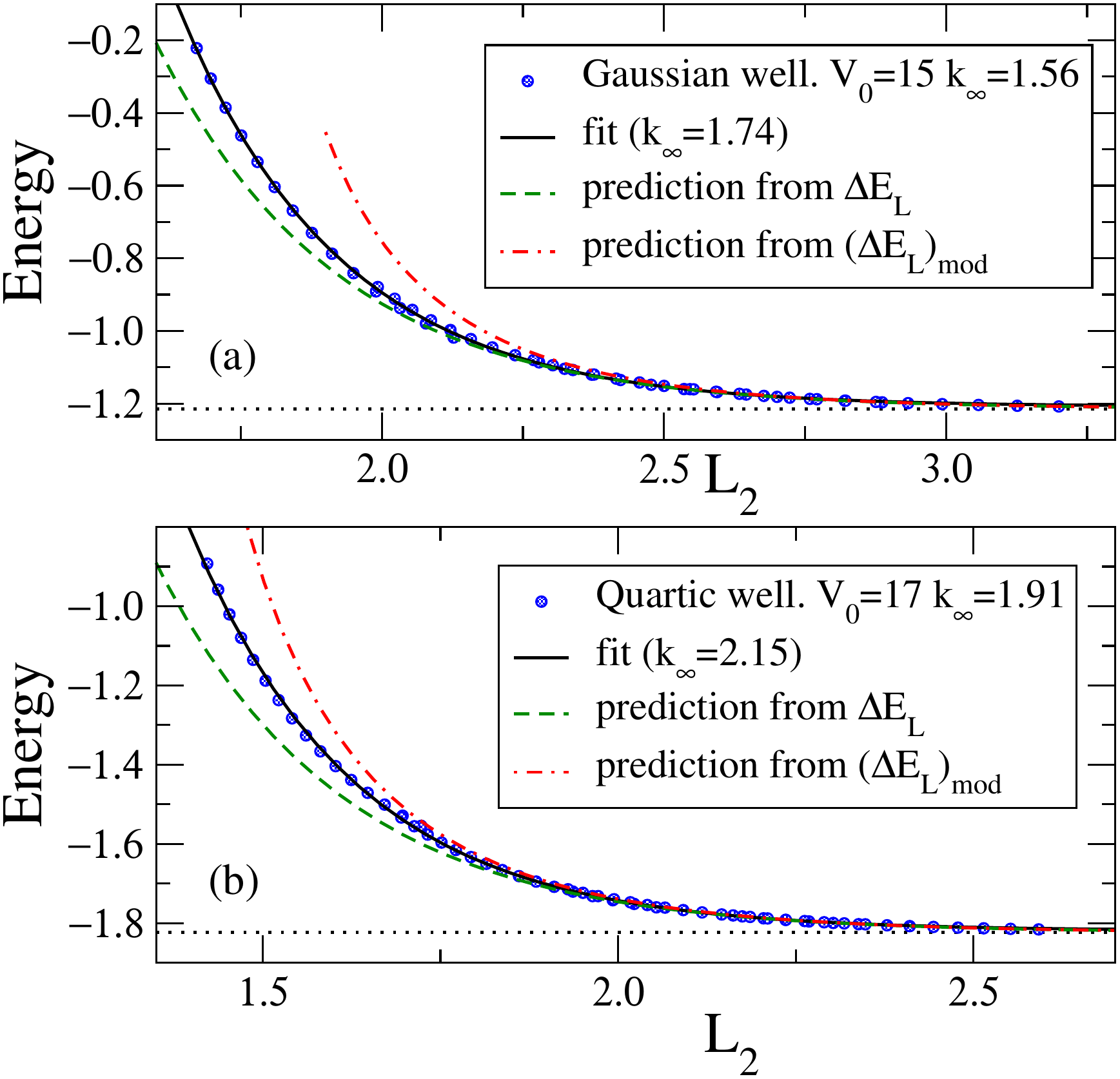}
\caption{(color online) Energy versus $L_2$ for the first excited
  states of deep (a) Gaussian Eq.~\eqref{eq:Vg} and (b) quartic
  Eq.~\eqref{eq:Vq} potential wells for a wide range of $N$ and $\hw$
  (circles) ($\hbar = \mu = R=1$).  The solid line is a fit to Eq.~\eqref{eq:IR_scaling1}
  with $A$, $\kinf$ and $\Einf$ as fit parameters while the dashed and
  dot-dashed lines are predictions from
  Eqs.~\eqref{eq:complete_IR_scaling} and Eq.~\eqref{eq:IR_scaling}.
  The horizontal dotted lines are the exact energies for the first excited states;
  Gaussian well: $\Einf = -1.2147$, quartic well: $\Einf = -1.8236$}
  \label{fig:IR_excited_gauss_quartic_wells}
\end{figure}

The results in Ref.~\cite{Furnstahl:2012qg} and the figures so far are
for the ground state of the potential.  However, the linear energy
approximation and the specific derivations in the last section should
also hold for excited states. This is so because the generalization of
the results in Subsection~\ref{subsec:analytic} shows that
$(j\pi/L_2)^2$ is a very good approximation to the $j^{\rm th}$
eigenvalue of the operator $p^2$ for $j \ll N$. In
Fig.~\ref{fig:IR_excited_gauss_quartic_wells} representative results
for excited states from two model potentials are shown.  We find the
same systematics as with the ground-state results: the exponential fit
works very well but the extracted $\kinf$ is only correct at about the 10\%
level.  In assessing the success of
Eqs.~\eqref{eq:complete_IR_scaling} and \eqref{eq:IR_scaling}, we note
that these excited states in deep potentials are comparable to the
ground states in moderate-depth potentials shown in
Fig.~\ref{fig:universal_sq_gauss_wells}.  The discussion there applies
here as well, namely that contributions from the second and third
terms in Eq.~\eqref{eq:entire_correction} are of comparable size and
opposite sign, so that Eq.~\eqref{eq:complete_IR_scaling} alone is a
better approximation.

In summary, our tests confirm the expectation from
Section~\ref{sec:universality} that the exponential form of
corrections for finite HO basis size is universal for different
potentials and also excited states (and also in one dimension, not shown).  
The
leading-order expression Eq.~\eqref{eq:IR_scaling1} is moderately
successful but not quantitative if exact values for $\kinf$ and $\ANC$
are used.  This implies that one should not expect to accurately
extract $\kinf$ from a fit to Eq.~\eqref{eq:IR_scaling1}.  The
modified energy correction Eq.~\eqref{eq:IR_scaling} is not an
improvement for deep potentials because it is not the dominant
subleading correction, but we expect it to be the most important
correction for shallow bound states (including the deuteron), which we
consider next.

\subsection{Shallow bound states}

\begin{figure}[tbh-]
 \includegraphics[width=0.95\columnwidth]{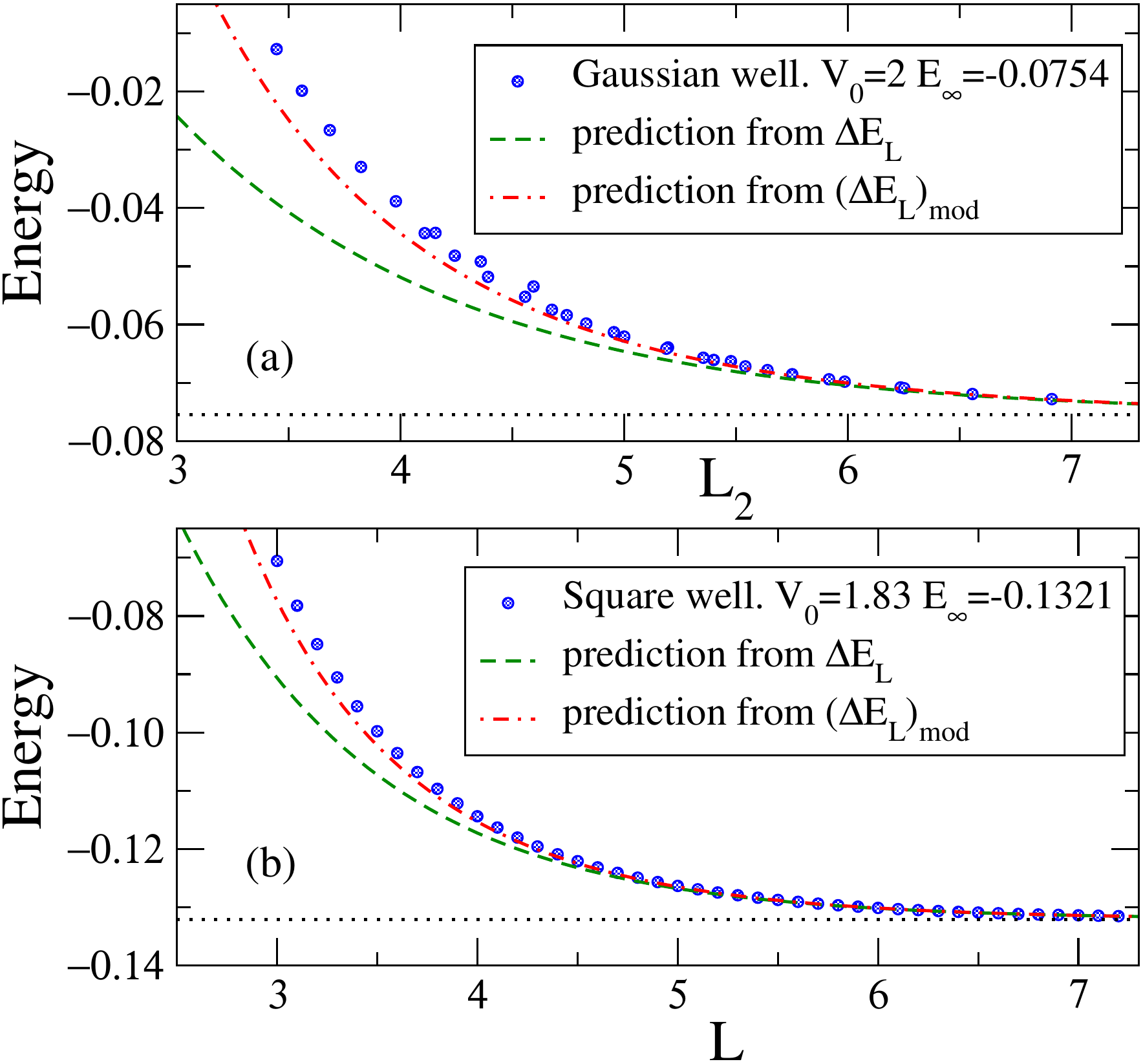}
 \caption{(color online) (a) Ground-state energy versus $L_2$ for 
   model Gaussian potential. (b) Energy versus $L$
   for the square well. The energies for the square well are from solving the Schr\"odinger equation
   exactly with a Dirichlet boundary condition on wave functions at
   $r=L$. The dashed and dot-dashed lines are predictions from
   Eqs.~\eqref{eq:complete_IR_scaling} and \eqref{eq:IR_scaling}. The depths of these model potentials
   are chosen so that the scaled
   energies (with $\hbar = \mu = R=1$) are the same as the deuteron binding energy.}
 \label{fig:mod_prediction_deuteron_equivalent}
\end{figure}

\begin{figure}[tbh-]
 \includegraphics[width=0.95\columnwidth]{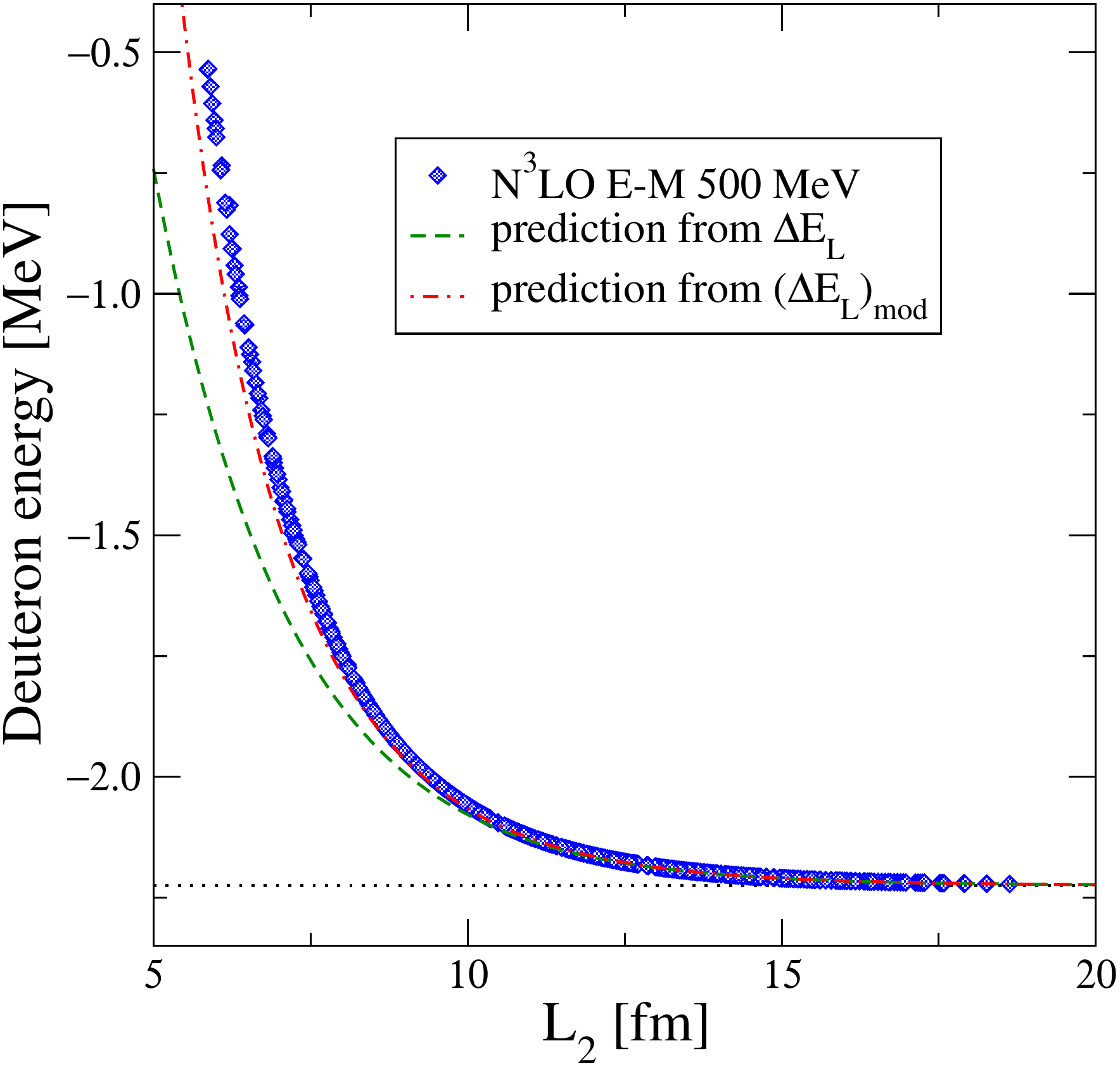}
 \caption{(color online) Deuteron energy versus $L_2$ for the
   potential of Ref.~\cite{Entem:2003ft}.  To eliminate the UV
   contamination we only keep points for which $\hw > 49$.  The dashed
   and dot-dashed lines are predictions from
   Eqs.~\eqref{eq:complete_IR_scaling} and \eqref{eq:IR_scaling}.  The
   horizontal dotted line is the deuteron binding energy.}
 \label{fig:EM_deuteron_data}
\end{figure}

The case of weakly bound states is of special interest.
Figure~\ref{fig:mod_prediction_deuteron_equivalent} (a) shows ground-state
energies for many different $N$ and $\hw$ versus $L_2$ using Gaussian model
potentials whose parameters are chosen so that the energies are the
same as the deuteron binding energy (scaled to units with $\hbar=1$,
$\mu=1$, $R=1$).  The prediction Eq.~\eqref{eq:complete_IR_scaling}
fails to reproduce the data except at the highest values of $L_2$.
However, when the correction from Eq.~\eqref{eq:IR_scaling} is added
there is significant improvement.  We also note that, contrary to the
situation with Figs.~\ref{fig:universal_sq_gauss_wells} and
\ref{fig:IR_excited_gauss_quartic_wells}, the correction from the
third term in Eq.~\eqref{eq:entire_correction} is much smaller and of
the same sign as the contribution from the second term included in
Eq.~\eqref{eq:IR_scaling}.  This is consistent with the dot-dashed
lines falling below the calculated energies at the smallest $L_2$
values. In Fig.~\ref{fig:mod_prediction_deuteron_equivalent} (b) the same 
exercise is repeated with a model square well. The energies in this case are obtained by
solving the Schr\"odinger equation exactly with a Dirichlet boundary condition on wave 
functions at $r=L$. Similar comments as for the model Gaussian potential well also apply here.

In Fig.~\ref{fig:EM_deuteron_data} we show analogous results from the
deuteron calculated with the chiral EFT potential of
Ref.~\cite{Entem:2003ft}.  As in
Fig.~\ref{fig:mod_prediction_deuteron_equivalent}, the modified IR
correction Eq.~\eqref{eq:IR_scaling} (evaluated using the $s$-wave
ANC) is a significant improvement over
Eq.~\eqref{eq:complete_IR_scaling}, falling slightly below the
calculations at the lowest $L_2$ values.

\subsection{Effect of SRG evolution}
  \label{subsec:SRG}

\begin{figure}
 \includegraphics[width=0.95\columnwidth]{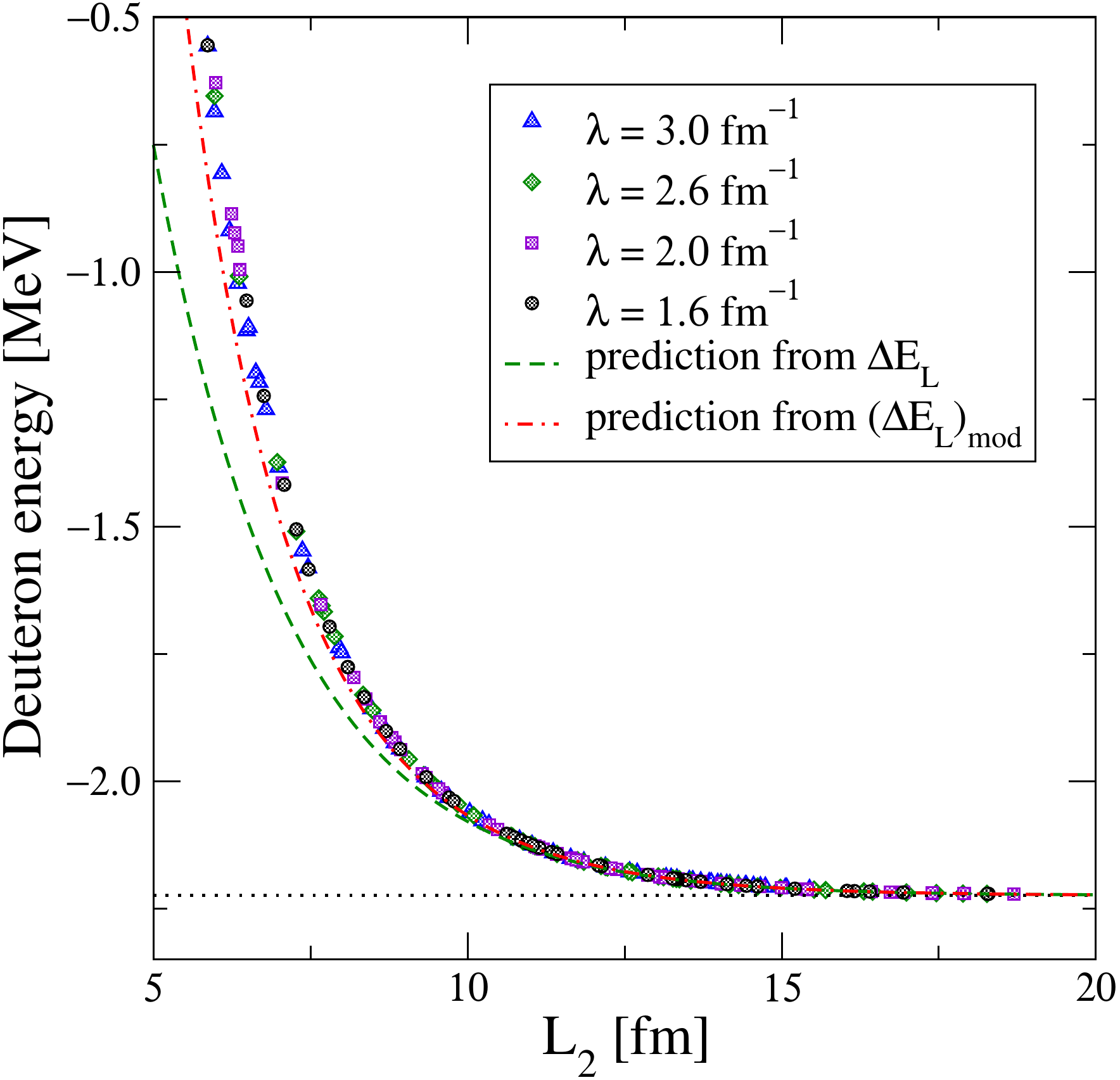}
 \caption{(color online) Deuteron energy versus $L_2$ for the
   potential of Ref.~\cite{Entem:2003ft} evolved by the SRG to four
   different resolutions (specified by $\lambda$).  To eliminate the
   UV contamination we only keep points for which $\hw >40$.  The
   dashed and dot-dashed lines are predictions from
   Eqs.~\eqref{eq:complete_IR_scaling} and \eqref{eq:IR_scaling}.  The
   horizontal dotted line is the deuteron binding energy.}
\label{fig:SRG_data_with_fit}
\end{figure}

\begin{figure}
 \includegraphics[width=0.95\columnwidth]{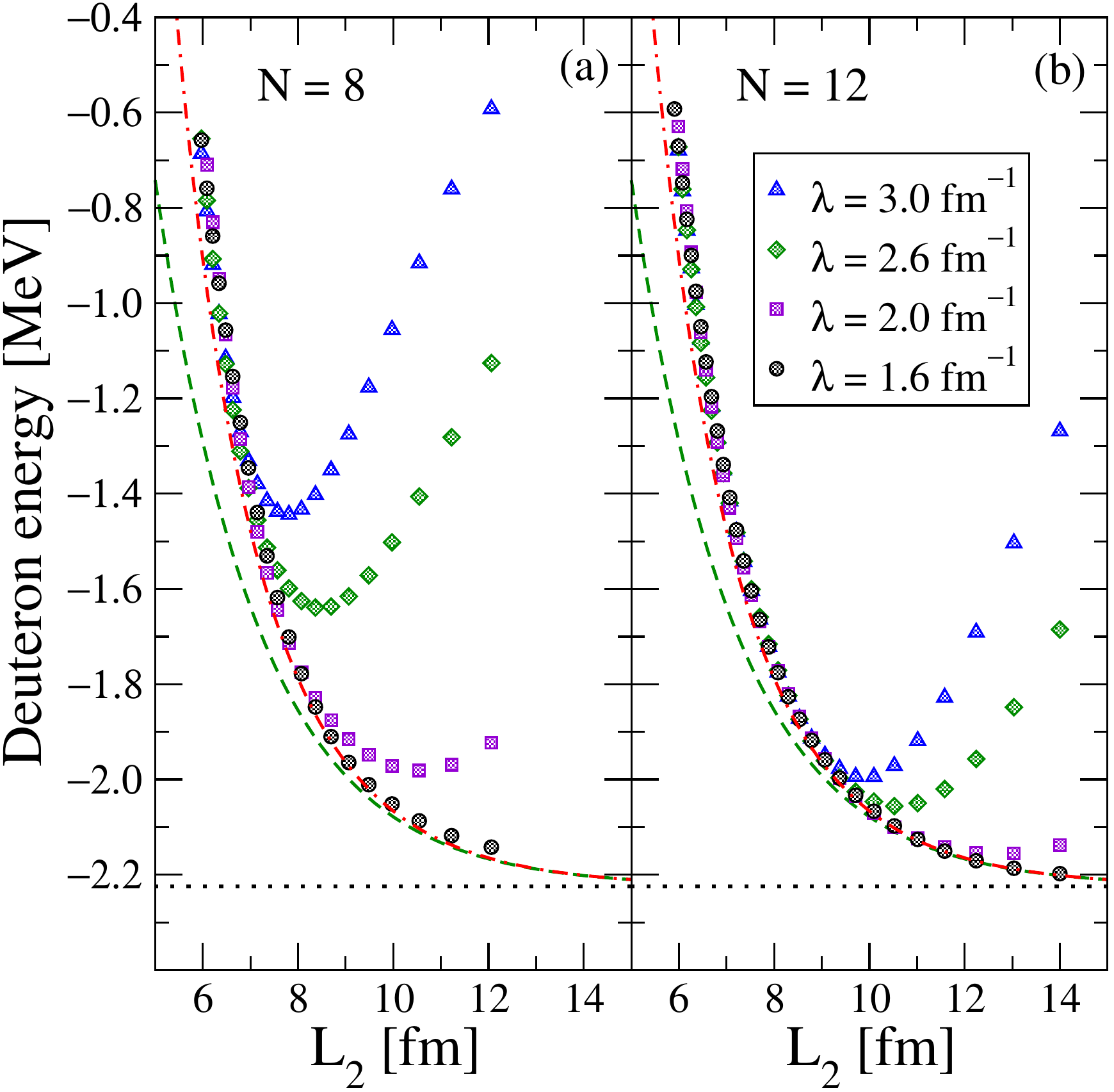}
 \caption{(color online) The same SRG-evolved potentials as in
   Fig.~\ref{fig:SRG_data_with_fit} are used to generate energies, but
   with $N$ fixed at (a) 8 and (b) 12 and no restriction on
   $\hw$.  Thus UV corrections are not negligible everywhere.  The
   dashed and dot-dashed lines are predictions from
   Eqs.~\eqref{eq:complete_IR_scaling} and \eqref{eq:IR_scaling}.  The
   horizontal dotted line is the deuteron binding energy.}
 \label{fig:SRG_data_collapse}
\end{figure}

As a final test of the universal applicability of the correction
formulas Eqs.~\eqref{eq:complete_IR_scaling} and
\eqref{eq:IR_scaling}, we consider a sequence of unitarily equivalent
potentials for the deuteron.  In particular, we use the similarity
renormalization group (SRG)~\cite{Bogner:2009bt} 
to evolve the initial Entem-Machleidt
potential to four values of the SRG evolution parameter $\lambda$.
Because the transformation is exactly unitary (up to very small
numerical errors) at the two-body level, the measurable quantities
such as phase shifts, bound-state energies, and ANCs are unchanged.
As $\lambda$ decreases, the SRG systematically reduces the coupling
between high-momentum and low-momentum potential matrix elements, thereby
lowering the effective UV cutoff.  Thus these potentials are useful
tools to assess the role of UV corrections.

We first consider results with $N$ and $\hw$ chosen to ensure small UV
corrections, as in all prior figures.  All the quantities on the RHS
of formula Eq.~\eqref{eq:IR_scaling} are invariant under SRG
evolution.  Therefore, if it is an accurate representation of the IR
energy corrections from truncating the HO basis, then the $E(L_2)$ vs
$L_2$ points for different SRG $\lambda$ should lie on the same curve.
Figure~\ref{fig:SRG_data_with_fit} shows that this is the case, 
and the curve is the same as for the unevolved potential in
Fig.~\ref{fig:EM_deuteron_data}.
(Only selected points are plotted for readability.)

Finally, in Fig.~\ref{fig:SRG_data_collapse} we relax the condition
that the UV corrections are small compared to IR corrections.  In
particular, we fix $N$ at 8 and 12 and scan through the full range of
$\hw$.  We observe that with increasing $L_2$, each of the curves with
a given $\lambda$ eventually deviates from the universal curve,
first with $\lambda = 3.0\fmi$ and then later with decreasing
$\lambda$ or with higher $N$.  We can understand this in terms of the
behavior of the induced UV cutoff.  For fixed $N$, Eq.~\eqref{eq:L_i}
tells us that increasing $L_2$ means increasing $b$ (or decreasing
$\hw$).  But at fixed $N$, $\LamUV \propto 1/b$, so the UV cutoff will
be decreasing and the corresponding UV energy correction increasing.
Thus the curves at fixed $\lambda$ correspond to the curves
seen in conventional plots of energy versus $\hw$ (e.g., see
Ref.~\cite{Bogner:2007rx}).  The softer potentials (lower $\lambda$)
will have lower intrinsic UV cutoffs and therefore they are only
affected for larger $L_2$. The minima for each $\lambda$ are when IR
and UV corrections are roughly equal.


\begin{figure}[t]
 \includegraphics[width=0.95\columnwidth]{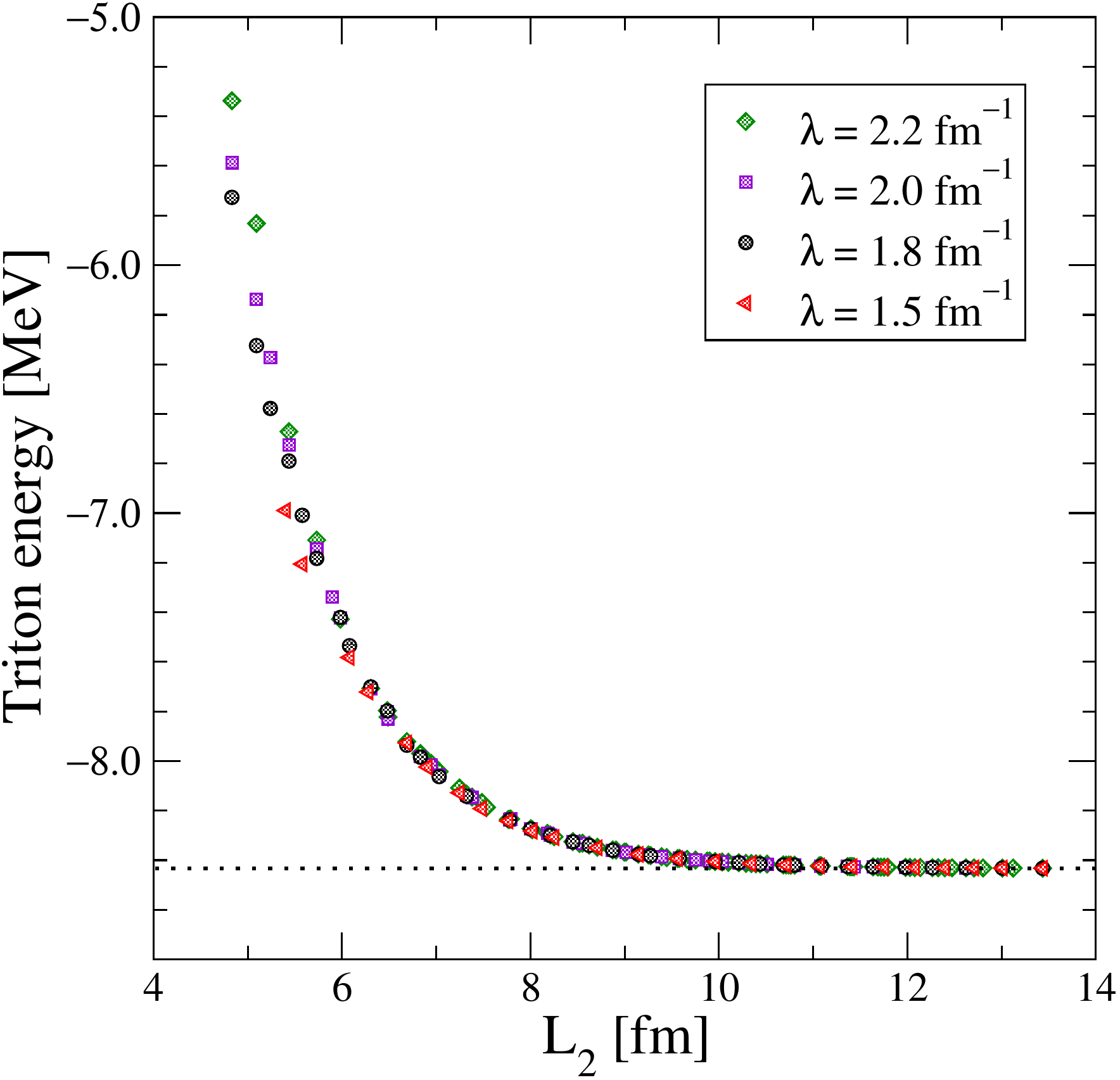}
 \caption{(color online) Triton energy versus $L_2$ (here
 calculated with the deuteron-neutron reduced mass) 
 for the two- and three-nucleon
 potential in Ref.~\cite{Jurgenson:2009qs} unitarily evolved by the SRG to four
   different resolutions (specified by $\lambda$) with the same
   binding energy~\cite{Jurgenson:2009qs,Jurgenson:2010wy}.  Only larger $\hw$ points are plotted
   to minimize the
   UV contamination.    The
   horizontal dotted line is the exact triton binding energy
   for this interaction.}
\label{fig:SRG_data_triton}
\end{figure}

\section{Summary and outlook}
\label{sec:summary}
  
In this paper, we revisited the infrared (IR) correction formula
derived in Ref.~\cite{Furnstahl:2012qg} for a truncated harmonic
oscillator (HO) basis expansion, using the simplified case of a
two-particle system as a controlled theoretical laboratory.  We used
simple model potentials and the deuteron calculated with realistic
potentials to extend and improve the IR formula. We demonstrated
analytically that the spectrum of the squared momentum operator $p^2$
in a finite oscillator basis is identical to the one in a spherical
box with a hard wall. The minimum eigenvalue of $p^2$ is
$(\pi\hbar/L_2)^2$, and this identifies $L_2$ as the box radius. While
these results have been obtained in finite but large oscillator
spaces, they also hold in practical applications in much smaller
spaces.  We showed how errors parametrized in terms of an effective
hard-wall radius $L$ from different $N$ and $\hw$ combinations all lie
on the same curve, but only if the UV error is sufficiently small and,
for smaller $N$, only if $L$ is defined as $L_2$ (see
Eq.~\eqref{eq:L_i}). The determination of $L_2$ as the box radius 
also allows us to extract phase shifts from the
positive-energy solutions in the oscillator basis.

The fall-off with $L_2$ of the IR correction to bound-state energies
is found to be an exponential independent of the potential
or whether a ground or excited states (or whether we are in one or
three dimensions).  This conclusion is validated by the derivation and
testing of explicit formulas for the energy corrections that depend
only on on measurable bound-state properties: the energy and residue
of the bound-state pole of the $S$ matrix (or the binding momentum and
asymptotic normalization constant).

Tests on larger nuclei have validated the exponential form
Eq.~\eqref{eq:IR_scaling1} with the decay parameter $\kinf$ in the
more general case associated with the lowest breakup threshold.
Preliminary tests show that $L_2$ is also the preferred definition of $L$.  
An example is shown in Fig.~\ref{fig:SRG_data_triton}, where triton
energies for a two- plus three-nucleon potential evolved
to four different SRG $\lambda$ 
(see Refs.~\cite{Jurgenson:2009qs,Jurgenson:2010wy})
lie on the same curve when $L_2$ is used.
A naive fit to Eq.~\eqref{eq:complete_IR_scaling} to the triton 
assuming a break-up into deuteron plus neutron
yields a binding momentum $\kinf\approx 91\,$MeV ($k_\infty^{\rm{expt}} = 88.13\,$MeV) and ANC
$\gamma_\infty\approx 3$~fm$^{-1/2}$. The ANC is not in agreement
with data and previous computations where $\gamma_\infty\approx 2$~fm$^{-1/2}$
was reported ~\cite{Huang:2008ye,Nollett:2011qf}, and suggests
that a more sophisticated analysis is necessary for the three-body
problem (see also Refs.~\cite{Kreuzer:2010ti,Polejaeva:2012ut,Briceno:2012rv}).
While we expect
from general considerations that the parameters
of universal curves such as in Fig.~\ref{fig:SRG_data_triton} 
are determined by
asymptotic (and therefore measurable) quantities, it remains to be
investigated whether simple formulas are possible (and whether ANC's
might be approximately extracted from fits).

In most of our investigations here we used our ability to calculate
with very large $\hw$ and $N$ for two-particle systems to ensure that
the effective UV cutoff was large enough to make the UV corrections
negligible compared to the IR corrections.  However, in realistic
calculations we will not (always) have this luxury.  The effects of
non-negligible UV corrections were shown in
Figure~\ref{fig:SRG_data_collapse}.  By working on the other side of
the minimum we can isolate the UV systematics.  Analogous studies to
those here but on the UV side show that $\LamUV$ is an appropriate
variable for the energy correction, 
but the behavior is \emph{not} universal in the same sense
we have identified here.  For example, considering different model
potentials, ground state vs.\ excited state, and three dimensions vs.\
one dimension, we find there are different functional dependencies
(see also Ref.~\cite{Tolle:2012cx}).  While some systematic behavior
has been identified for SRG-evolved
potentials~\cite{Furnstahl:2012qg}, further work is needed to go
beyond the basic form used to make fits.  Work in this direction is in
progress.

\medskip

\begin{acknowledgments}
  We thank R. Brice\~no, A. Bulgac, Z. Davoudi, K. Hebeler, H. Hergert, R. Perry, and K.  Wendt
  for useful discussions, K. Wendt for
  generating deuteron eigenvalues with SRG-evolved potentials for a
  very wide range of \hw\ and $N$, and E. Jurgenson for triton
  results.  This work was supported in part by the National Science
  Foundation under Grant No.~PHY--1002478 and the Department of Energy
  under Grant Nos.~DE-FG02-96ER40963 (University of Tennessee),
  DE-AC05-00OR22725 (Oak Ridge National Laboratory), and
  DE-SC0008499/DE-SC0008533 (SciDAC-3 NUCLEI project), and by the
  Swedish Research Council.
\end{acknowledgments}


\begin{thebibliography}{34}
\expandafter\ifx\csname natexlab\endcsname\relax\def\natexlab#1{#1}\fi
\expandafter\ifx\csname bibnamefont\endcsname\relax
  \def\bibnamefont#1{#1}\fi
\expandafter\ifx\csname bibfnamefont\endcsname\relax
  \def\bibfnamefont#1{#1}\fi
\expandafter\ifx\csname citenamefont\endcsname\relax
  \def\citenamefont#1{#1}\fi
\expandafter\ifx\csname url\endcsname\relax
  \def\url#1{\texttt{#1}}\fi
\expandafter\ifx\csname urlprefix\endcsname\relax\def\urlprefix{URL }\fi
\providecommand{\bibinfo}[2]{#2}
\providecommand{\eprint}[2][]{\url{#2}}

\bibitem[{\citenamefont{Coon et~al.}(2012)\citenamefont{Coon, Avetian, Kruse,
  van Kolck, Maris et~al.}}]{Coon:2012ab}
\bibinfo{author}{\bibfnamefont{S.~A.} \bibnamefont{Coon}},
  \bibinfo{author}{\bibfnamefont{M.~I.} \bibnamefont{Avetian}},
  \bibinfo{author}{\bibfnamefont{M.~K.} \bibnamefont{Kruse}},
  \bibinfo{author}{\bibfnamefont{U.}~\bibnamefont{van Kolck}},
  \bibinfo{author}{\bibfnamefont{P.}~\bibnamefont{Maris}},
  \bibnamefont{et~al.}, \bibinfo{journal}{Phys. Rev. C}
  \textbf{\bibinfo{volume}{86}}, \bibinfo{pages}{054002}
  (\bibinfo{year}{2012}).

\bibitem[{\citenamefont{Furnstahl et~al.}(2012)\citenamefont{Furnstahl, Hagen,
  and Papenbrock}}]{Furnstahl:2012qg}
\bibinfo{author}{\bibfnamefont{R.}~\bibnamefont{Furnstahl}},
  \bibinfo{author}{\bibfnamefont{G.}~\bibnamefont{Hagen}}, \bibnamefont{and}
  \bibinfo{author}{\bibfnamefont{T.}~\bibnamefont{Papenbrock}},
  \bibinfo{journal}{Phys. Rev. C} \textbf{\bibinfo{volume}{86}},
  \bibinfo{pages}{031301} (\bibinfo{year}{2012}).

\bibitem[{\citenamefont{L{\"u}scher}(1986)}]{Luscher:1985dn}
\bibinfo{author}{\bibfnamefont{M.}~\bibnamefont{L{\"u}scher}},
  \bibinfo{journal}{Commun. Math. Phys.} \textbf{\bibinfo{volume}{104}},
  \bibinfo{pages}{177} (\bibinfo{year}{1986}).

\bibitem[{\citenamefont{Lee and Pine}(2011)}]{Lee:2010km}
\bibinfo{author}{\bibfnamefont{D.}~\bibnamefont{Lee}} \bibnamefont{and}
  \bibinfo{author}{\bibfnamefont{M.}~\bibnamefont{Pine}},
  \bibinfo{journal}{Eur. Phys. J. A} \textbf{\bibinfo{volume}{47}},
  \bibinfo{pages}{41} (\bibinfo{year}{2011}).

\bibitem[{\citenamefont{Pine and Lee}(2012)}]{Pine:2012zv}
\bibinfo{author}{\bibfnamefont{M.}~\bibnamefont{Pine}} \bibnamefont{and}
  \bibinfo{author}{\bibfnamefont{D.}~\bibnamefont{Lee}}, \bibinfo{journal}{Annals
  Phys.} \textbf{\bibinfo{volume}{331}},
  \bibinfo{pages}{24} (\bibinfo{year}{2013}).


\bibitem[{\citenamefont{Koenig et~al.}(2012)\citenamefont{Koenig, Lee, and
  Hammer}}]{Konig:2011ti}
\bibinfo{author}{\bibfnamefont{S.}~\bibnamefont{Koenig}},
  \bibinfo{author}{\bibfnamefont{D.}~\bibnamefont{Lee}}, \bibnamefont{and}
  \bibinfo{author}{\bibfnamefont{H.-W.} \bibnamefont{Hammer}},
  \bibinfo{journal}{Annals Phys.} \textbf{\bibinfo{volume}{327}},
  \bibinfo{pages}{1450} (\bibinfo{year}{2012}).

\bibitem[{\citenamefont{Entem and Machleidt}(2003)}]{Entem:2003ft}
\bibinfo{author}{\bibfnamefont{D.~R.} \bibnamefont{Entem}} \bibnamefont{and}
  \bibinfo{author}{\bibfnamefont{R.}~\bibnamefont{Machleidt}},
  \bibinfo{journal}{Phys. Rev. C} \textbf{\bibinfo{volume}{68}},
  \bibinfo{pages}{041001} (\bibinfo{year}{2003}).

\bibitem[{\citenamefont{Bogner et~al.}(2008)\citenamefont{Bogner, Furnstahl,
  Maris, Perry, Schwenk, and Vary}}]{Bogner:2007rx}
\bibinfo{author}{\bibfnamefont{S.~K.} \bibnamefont{Bogner}},
  \bibinfo{author}{\bibfnamefont{R.~J.} \bibnamefont{Furnstahl}},
  \bibinfo{author}{\bibfnamefont{P.}~\bibnamefont{Maris}},
  \bibinfo{author}{\bibfnamefont{R.~J.} \bibnamefont{Perry}},
  \bibinfo{author}{\bibfnamefont{A.}~\bibnamefont{Schwenk}}, \bibnamefont{and}
  \bibinfo{author}{\bibfnamefont{J.~P.} \bibnamefont{Vary}},
  \bibinfo{journal}{Nucl. Phys. A} \textbf{\bibinfo{volume}{801}},
  \bibinfo{pages}{21} (\bibinfo{year}{2008}).

\bibitem[{\citenamefont{Gradshteyn and Ryzhik}(2000)}]{gradshteyn}
\bibinfo{author}{\bibfnamefont{L.~S.} \bibnamefont{Gradshteyn}}
  \bibnamefont{and} \bibinfo{author}{\bibfnamefont{L.~M.}
  \bibnamefont{Ryzhik}}, \emph{\bibinfo{title}{Tables of integrals, series, and
  products}} (\bibinfo{publisher}{Academic Press}, \bibinfo{address}{San
  Diego}, \bibinfo{year}{2000}), \bibinfo{edition}{6th} ed.

\bibitem[{\citenamefont{Stetcu et~al.}(2007{\natexlab{a}})\citenamefont{Stetcu,
  Barrett, and van Kolck}}]{Stetcu:2006ey}
\bibinfo{author}{\bibfnamefont{I.}~\bibnamefont{Stetcu}},
  \bibinfo{author}{\bibfnamefont{B.~R.} \bibnamefont{Barrett}},
  \bibnamefont{and} \bibinfo{author}{\bibfnamefont{U.}~\bibnamefont{van
  Kolck}}, \bibinfo{journal}{Phys. Lett. B} \textbf{\bibinfo{volume}{653}},
  \bibinfo{pages}{358} (\bibinfo{year}{2007}{\natexlab{a}}).


\bibitem[{\citenamefont{{Bang} et~al.}(2000)\citenamefont{{Bang}, {Mazur},
  {Shirokov}, {Smirnov}, and {Zaytsev}}}]{bang2000}
\bibinfo{author}{\bibfnamefont{J.~M.} \bibnamefont{{Bang}}},
  \bibinfo{author}{\bibfnamefont{A.~I.} \bibnamefont{{Mazur}}},
  \bibinfo{author}{\bibfnamefont{A.~M.} \bibnamefont{{Shirokov}}},
  \bibinfo{author}{\bibfnamefont{Y.~F.} \bibnamefont{{Smirnov}}},
  \bibnamefont{and} \bibinfo{author}{\bibfnamefont{S.~A.}
  \bibnamefont{{Zaytsev}}}, \bibinfo{journal}{Annals Phys.}
  \textbf{\bibinfo{volume}{280}}, \bibinfo{pages}{299} (\bibinfo{year}{2000}).

\bibitem[{\citenamefont{{Luu} et~al.}(2010)\citenamefont{{Luu}, {Savage},
  {Schwenk}, and {Vary}}}]{Luu2010}
\bibinfo{author}{\bibfnamefont{T.}~\bibnamefont{{Luu}}},
  \bibinfo{author}{\bibfnamefont{M.~J.} \bibnamefont{{Savage}}},
  \bibinfo{author}{\bibfnamefont{A.}~\bibnamefont{{Schwenk}}},
  \bibnamefont{and} \bibinfo{author}{\bibfnamefont{J.~P.}
  \bibnamefont{{Vary}}}, \bibinfo{journal}{\prc} \textbf{\bibinfo{volume}{82}},
  \bibinfo{eid}{034003} (\bibinfo{year}{2010}).

\bibitem[{\citenamefont{Stetcu et~al.}(2010)\citenamefont{Stetcu, Rotureau,
  Barrett, and van Kolck}}]{Stetcu2010jpg}
\bibinfo{author}{\bibfnamefont{I.}~\bibnamefont{Stetcu}},
  \bibinfo{author}{\bibfnamefont{J.}~\bibnamefont{Rotureau}},
  \bibinfo{author}{\bibfnamefont{B.~R.} \bibnamefont{Barrett}},
  \bibnamefont{and} \bibinfo{author}{\bibfnamefont{U.}~\bibnamefont{van
  Kolck}}, \bibinfo{journal}{Journal of Physics G: Nuclear and Particle
  Physics} \textbf{\bibinfo{volume}{37}}, \bibinfo{pages}{064033}
  (\bibinfo{year}{2010}).

\bibitem[{\citenamefont{Busch et~al.}(1998)\citenamefont{Busch, Englert,
  Rzazewski, and Wilkens}}]{busch1998}
\bibinfo{author}{\bibfnamefont{T.}~\bibnamefont{Busch}},
  \bibinfo{author}{\bibfnamefont{B.-G.} \bibnamefont{Englert}},
  \bibinfo{author}{\bibfnamefont{K.}~\bibnamefont{Rzazewski}},
  \bibnamefont{and} \bibinfo{author}{\bibfnamefont{M.}~\bibnamefont{Wilkens}},
  \bibinfo{journal}{Foundations of Physics} \textbf{\bibinfo{volume}{28}},
  \bibinfo{pages}{549} (\bibinfo{year}{1998}).

\bibitem[{\citenamefont{{Bhattacharyya} and
  {Papenbrock}}(2006)}]{Bhattacharyya2006}
\bibinfo{author}{\bibfnamefont{A.}~\bibnamefont{{Bhattacharyya}}}
  \bibnamefont{and}
  \bibinfo{author}{\bibfnamefont{T.}~\bibnamefont{{Papenbrock}}},
  \bibinfo{journal}{\pra} \textbf{\bibinfo{volume}{74}}, \bibinfo{eid}{041602}
  (\bibinfo{year}{2006}).

\bibitem[{\citenamefont{Hagen et~al.}(2007)\citenamefont{Hagen, Dean,
  Hjorth-Jensen, Papenbrock, and Schwenk}}]{Hagen:2007hi}
\bibinfo{author}{\bibfnamefont{G.}~\bibnamefont{Hagen}},
  \bibinfo{author}{\bibfnamefont{D.~J.} \bibnamefont{Dean}},
  \bibinfo{author}{\bibfnamefont{M.}~\bibnamefont{Hjorth-Jensen}},
  \bibinfo{author}{\bibfnamefont{T.}~\bibnamefont{Papenbrock}},
  \bibnamefont{and} \bibinfo{author}{\bibfnamefont{A.}~\bibnamefont{Schwenk}},
  \bibinfo{journal}{Phys. Rev. C} \textbf{\bibinfo{volume}{76}},
  \bibinfo{pages}{044305} (\bibinfo{year}{2007}).

\bibitem[{\citenamefont{Forssen et~al.}(2008)\citenamefont{Forssen, Vary,
  Caurier, and Navratil}}]{Forssen:2008qp}
\bibinfo{author}{\bibfnamefont{C.}~\bibnamefont{Forssen}},
  \bibinfo{author}{\bibfnamefont{J.}~\bibnamefont{Vary}},
  \bibinfo{author}{\bibfnamefont{E.}~\bibnamefont{Caurier}}, \bibnamefont{and}
  \bibinfo{author}{\bibfnamefont{P.}~\bibnamefont{Navratil}},
  \bibinfo{journal}{Phys. Rev. C} \textbf{\bibinfo{volume}{77}},
  \bibinfo{pages}{024301} (\bibinfo{year}{2008}).

\bibitem[{\citenamefont{Maris et~al.}(2009)\citenamefont{Maris, Vary, and
  Shirokov}}]{Maris:2008ax}
\bibinfo{author}{\bibfnamefont{P.}~\bibnamefont{Maris}},
  \bibinfo{author}{\bibfnamefont{J.~P.} \bibnamefont{Vary}}, \bibnamefont{and}
  \bibinfo{author}{\bibfnamefont{A.~M.} \bibnamefont{Shirokov}},
  \bibinfo{journal}{Phys. Rev. C} \textbf{\bibinfo{volume}{79}},
  \bibinfo{pages}{014308} (\bibinfo{year}{2009}).

\bibitem[{\citenamefont{Roth}(2009)}]{Roth:2009cw}
\bibinfo{author}{\bibfnamefont{R.}~\bibnamefont{Roth}}, \bibinfo{journal}{Phys.
  Rev. C} \textbf{\bibinfo{volume}{79}}, \bibinfo{pages}{064324}
  (\bibinfo{year}{2009}).

\bibitem[{\citenamefont{Tolle et~al.}(2012)\citenamefont{Tolle, Hammer, and
  Metsch}}]{Tolle:2012cx}
\bibinfo{author}{\bibfnamefont{S.}~\bibnamefont{T\"{o}lle}},
  \bibinfo{author}{\bibfnamefont{H.-W.} \bibnamefont{Hammer}},
  \bibnamefont{and} \bibinfo{author}{\bibfnamefont{B.~Ch.} \bibnamefont{Metsch}},
  \bibinfo{journal}{J. Phys. G: Nucl. Part. Phys.} \textbf{\bibinfo{volume}{40}},
  \bibinfo{pages}{055004} (\bibinfo{year}{2013}).

\bibitem[{\citenamefont{Soma et~al.}(2013)\citenamefont{Soma, Barbieri, and
  Duguet}}]{Soma:2012zd}
\bibinfo{author}{\bibfnamefont{V.}~\bibnamefont{Soma}},
  \bibinfo{author}{\bibfnamefont{C.}~\bibnamefont{Barbieri}}, \bibnamefont{and}
  \bibinfo{author}{\bibfnamefont{T.}~\bibnamefont{Duguet}},
  \bibinfo{journal}{Phys. Rev. C} \textbf{\bibinfo{volume}{87}},
  \bibinfo{pages}{011303} (\bibinfo{year}{2013}).

\bibitem[{\citenamefont{Hergert et~al.}(2012)\citenamefont{Hergert, Bogner,
  Binder, Calci, Langhammer et~al.}}]{Hergert:2012nb}
\bibinfo{author}{\bibfnamefont{H.}~\bibnamefont{Hergert}},
  \bibinfo{author}{\bibfnamefont{S.}~\bibnamefont{Bogner}},
  \bibinfo{author}{\bibfnamefont{S.}~\bibnamefont{Binder}},
  \bibinfo{author}{\bibfnamefont{A.}~\bibnamefont{Calci}},
  \bibinfo{author}{\bibfnamefont{J.}~\bibnamefont{Langhammer}},
  \bibnamefont{et~al.}, \bibinfo{journal}{Phys. Rev. C} \textbf{\bibinfo{volume}{87}},
  \bibinfo{pages}{034307} (\bibinfo{year}{2013}).


\bibitem[{\citenamefont{Djajaputra and Cooper}(2000)}]{Djajaputra:2000aa}
\bibinfo{author}{\bibfnamefont{D.}~\bibnamefont{Djajaputra}} \bibnamefont{and}
  \bibinfo{author}{\bibfnamefont{B.~R.} \bibnamefont{Cooper}},
  \bibinfo{journal}{European Journal of Physics} \textbf{\bibinfo{volume}{21}},
  \bibinfo{pages}{261} (\bibinfo{year}{2000}).

\bibitem[{\citenamefont{Taylor}(2006)}]{taylor2006scattering}
\bibinfo{author}{\bibfnamefont{J.}~\bibnamefont{Taylor}},
  \emph{\bibinfo{title}{Scattering Theory: The Quantum Theory of
  Nonrelativistic Collisions}} (\bibinfo{publisher}{Dover},
  \bibinfo{year}{2006}).

\bibitem[{\citenamefont{Amado}(1979)}]{Amado:1979zz}
\bibinfo{author}{\bibfnamefont{R.~D.} \bibnamefont{Amado}},
  \bibinfo{journal}{Phys. Rev. C} \textbf{\bibinfo{volume}{19}},
  \bibinfo{pages}{1473} (\bibinfo{year}{1979}).

\bibitem[{\citenamefont{Bogner et~al.}(2010)\citenamefont{Bogner, Furnstahl,
  and Schwenk}}]{Bogner:2009bt}
\bibinfo{author}{\bibfnamefont{S.~K.} \bibnamefont{Bogner}},
  \bibinfo{author}{\bibfnamefont{R.~J.} \bibnamefont{Furnstahl}},
  \bibnamefont{and} \bibinfo{author}{\bibfnamefont{A.}~\bibnamefont{Schwenk}},
  \bibinfo{journal}{Prog. Part. Nucl. Phys.} \textbf{\bibinfo{volume}{65}},
  \bibinfo{pages}{94} (\bibinfo{year}{2010}).

\bibitem[{\citenamefont{Jurgenson et~al.}(2009)\citenamefont{Jurgenson,
  Navratil, and Furnstahl}}]{Jurgenson:2009qs}
\bibinfo{author}{\bibfnamefont{E.~D.} \bibnamefont{Jurgenson}},
  \bibinfo{author}{\bibfnamefont{P.}~\bibnamefont{Navratil}}, \bibnamefont{and}
  \bibinfo{author}{\bibfnamefont{R.~J.} \bibnamefont{Furnstahl}},
  \bibinfo{journal}{Phys. Rev. Lett.} \textbf{\bibinfo{volume}{103}},
  \bibinfo{pages}{082501} (\bibinfo{year}{2009}).

\bibitem[{\citenamefont{Jurgenson et~al.}(2011)\citenamefont{Jurgenson,
  Navratil, and Furnstahl}}]{Jurgenson:2010wy}
\bibinfo{author}{\bibfnamefont{E.~D.} \bibnamefont{Jurgenson}},
  \bibinfo{author}{\bibfnamefont{P.}~\bibnamefont{Navratil}}, \bibnamefont{and}
  \bibinfo{author}{\bibfnamefont{R.~J.} \bibnamefont{Furnstahl}},
  \bibinfo{journal}{Phys. Rev. C} \textbf{\bibinfo{volume}{83}},
  \bibinfo{pages}{034301} (\bibinfo{year}{2011}).

\bibitem[{\citenamefont{Huang et~al.}(2010)\citenamefont{Huang, Bertulani, and
  Guimaraes}}]{Huang:2008ye}
\bibinfo{author}{\bibfnamefont{J.}~\bibnamefont{Huang}},
  \bibinfo{author}{\bibfnamefont{C.}~\bibnamefont{Bertulani}},
  \bibnamefont{and}
  \bibinfo{author}{\bibfnamefont{V.}~\bibnamefont{Guimaraes}},
  \bibinfo{journal}{Atom. Data Nucl. Data Tabl.} \textbf{\bibinfo{volume}{96}},
  \bibinfo{pages}{824} (\bibinfo{year}{2010}).

\bibitem[{\citenamefont{Nollett and Wiringa}(2011)}]{Nollett:2011qf}
\bibinfo{author}{\bibfnamefont{K.~M.} \bibnamefont{Nollett}} \bibnamefont{and}
  \bibinfo{author}{\bibfnamefont{R.}~\bibnamefont{Wiringa}},
  \bibinfo{journal}{Phys. Rev. C} \textbf{\bibinfo{volume}{83}},
  \bibinfo{pages}{041001} (\bibinfo{year}{2011}).

\bibitem[{\citenamefont{Kreuzer and Hammer}(2011)}]{Kreuzer:2010ti}
\bibinfo{author}{\bibfnamefont{S.}~\bibnamefont{Kreuzer}} \bibnamefont{and}
  \bibinfo{author}{\bibfnamefont{H.-W.} \bibnamefont{Hammer}},
  \bibinfo{journal}{Phys. Lett.} \textbf{\bibinfo{volume}{B694}},
  \bibinfo{pages}{424} (\bibinfo{year}{2011}).

\bibitem[{\citenamefont{Polejaeva and Rusetsky}(2012)}]{Polejaeva:2012ut}
\bibinfo{author}{\bibfnamefont{K.}~\bibnamefont{Polejaeva}} \bibnamefont{and}
  \bibinfo{author}{\bibfnamefont{A.}~\bibnamefont{Rusetsky}},
  \bibinfo{journal}{Eur. Phys. J.} \textbf{\bibinfo{volume}{A48}},
  \bibinfo{pages}{67} (\bibinfo{year}{2012}).

\bibitem[{\citenamefont{Briceno and Davoudi}(2012)}]{Briceno:2012rv}
\bibinfo{author}{\bibfnamefont{R.~A.} \bibnamefont{Briceno}} \bibnamefont{and}
  \bibinfo{author}{\bibfnamefont{Z.}~\bibnamefont{Davoudi}}
  (\bibinfo{year}{2012}), \eprint{arXiv:1212.3398}.

\end{thebibliography}

\end{document}